\newcounter{myctr}
\def\myitem{\refstepcounter{myctr}\bibfont\noindent\ifnum\themyctr>9\else\phantom{0}\fi\hangindent17pt\themyctr.\enskip}

\documentclass{ws-ijqi}
\usepackage{hyperref}
\usepackage[super,sort,compress]{cite}

\newcommand{\ket}[1]{\ensuremath{\left|#1\right\rangle}}

\newcommand{\uspin}{\ket{\uparrow}}
\newcommand{\dspin}{\ket{\downarrow}}

\newcommand{\uspinbar}{\ket{\bar{\uparrow}}}
\newcommand{\dspinbar}{\ket{\bar{\downarrow}}}
\newcommand{\psibar}[1]{\bar{\Psi}^{\bar{#1}}}
\newcommand{\phibar}[1]{\bar{\Phi}^{\bar{#1}}}
\newcommand{\oneHalf}{{1 \over 2}}

\begin{document}

\catchline{}{}{}{}{}

\title{The Frauchiger-Renner Gedanken Experiment:\\
Flaws in Its Analysis -- \\ 
How Logic Works in Quantum Mechanics\footnote{A theory seminar 
presenting some of the results of this paper is available at 
\newline 
\hspace*{3 mm} https://youtu.be/ODP6cUj8vxY (Accessed on 9/10/23).}}

\author{Stuart Samuel\footnote{Retired Professor 
\newline
\hspace*{1.5 mm}Previous Institutions (in historical order):  
\newline
\hspace*{5 mm} Institute for Advanced Study (Princeton, New Jersey)
\newline
\hspace*{5 mm} Columbia University (New York, NY)
\newline
\hspace*{5 mm} City College of New York
\newline
E-mail: stuartsamuel@hotmail.com}}

\maketitle


\begin{abstract}

In a publication (Nature Comm. 3711, 9 (2018)), 
Daniela Frauchiger and Renato Renner used 
a Wigner/friend gedanken experiment to argue 
that quantum mechanics cannot describe complex 
systems involving measuring agents. 
The result was framed as a theorem: 
Three assumptions about quantum mechanics were made 
and a ``proof'' was then produced. 
However, Frauchiger and Renner overlooked three other important assumptions 
that were implicitly used in the derivation of their theorem, 
two of which separately nullify their result. 
Basically, Frauchiger and Renner applied ``classical'' logic and thinking 
to a situation that requires ``quantum'' logic. 
There are also other problems with the analysis in the publication, 
which are not fatal but have consequences for certain aspects of their work. 
Their gedanken experiment does not show that quantum mechanics 
cannot be applied to certain macroscopic systems; 
instead, it demonstrates that the transitive property of logic 
fails in certain quantum mechanical situations. 

We find the gedanken experiment useful in the sense that it has prompted us 
to investigate how logic works for statements about wavefunctions 
and measurements in quantum mechanics.

\vspace{5mm} 
\end{abstract}


\markboth{Stuart Samuel}
{The Frauchiger-Renner Gedanken Experiment in Unitary Quantum Mechanics}


\section{Introduction}
\label{intro}

In a publication,\cite{FRGE} D. Frauchiger and R. Renner argued  
that quantum theory cannot consistently describe the use of itself 
in the following sense: 
If quantum mechanics governs the agents involved in experiments,  
then a gedanken experiment exists that leads to a contradiction. 
By "agents," one means  the experimentalists and their equipment 
that record the results of measurements on quantum systems. 
An experimentalist is able to note the outcome of an experiment  
through statements such as ``I observed the spin of a spin-$\oneHalf$ object to be up" 
(or down if that is the result of the measurement) where ``up'' and ``down'' refer to the direction of the spin in, say, in the $z$-direction. 
The equipment of the experimentalist equally has this capacity 
by recording the result (be it up or down) in a database, for example. 

In the Frauchiger-Renner gedanken experiment, 
the initial state involves two spin-$\oneHalf$ objects that are entangled. 
Four measurements are performed: 
In the first two, two different agents measure the spin of each object in the $z$-direction. 
In the final two, Wigner agents perform measurements on those two agents themselves,\cite{WignersFriendA,WignersFriendB} 
and therefore, quantum mechanically measure macroscopic systems. 
Four ``If ..., then ...'' statements concerning the measurements are derived. 
These statements are then combined using the transitive 
property of logic (if $A$ implies $B$ and $B$ implies $C$, then $A$ implies $C$) 
three times to arrive at a contradiction.
From this, Frauchiger and Renner conclude that ``quantum theory cannot be
extrapolated to complex systems, at least not in a straightforward manner.''

The two authors cast the above result as a theorem involving three assumptions about quantum mechanics. 
We refer the reader to the details of these assumptions in reference \refcite{FRGE}, 
and, instead, in this Introduction, describe them in general terms. 
The first assumption is the ``quantum mechanical'' one (Q). 
It says that the probability of an output of a measurement is given by the Born rule. 
The second assumption (C) demands consistency,  
meaning that if one agent reaches a conclusion about a statement or prediction, 
then another agent, when using the same theory, assumptions and information, 
will arrive to the same conclusion. 
Assumption (S) asserts that if an agent is certain of the outcome of a measurement or statement, 
then the agent cannot conclude something contradicting it. 
Assumption (S) may appear to be obvious, 
but the Contradictory Statement (see Section \ref{TheArgument}) that arises in the Frauchiger-Renner gedanken experiment 
takes the form ``If I (Agent W) obtain a measurement of  `OK', then by using assumptions (Q) and (C), 
I can conclude that I should obtain a measurement of `FAIL'.
Here, `OK' and  `FAIL' are similar to the down and up states of a spin-$\oneHalf$ object 
when the axis of spin quantization is in the $x$-direction. 
So, the argument in the gedanken experiment makes use of (S). 

Frauchiger and Renner leave open the possibility that there are implicit assumptions beyond the above three: 
``Any no-go result ... is phrased within a particular framework that comes
with a set of built-in assumptions. 
Hence, it is always possible that a theory evades the conclusions of the no-go result by not fulfilling
these implicit assumptions.''\cite{FRGE}  
Indeed, there are many overlooked assumptions. 

The first omitted assumption in the Frauchiger and Renner paper is that wavefunction collapse does not occur. 
If wavefunction collapse occurs during a measurement, then one or more of the four ``If ..., then ...'' statements is rendered false 
and the contradiction cannot be generated. 
This is discussed at the end of Section \ref{TheArgument}, 
but intuitively, it can be understood as follows. 
Three of the four ``If ..., then ...'' statements sensitively depend on cancellations 
between different terms in the wavefunction. 
The slightest change in the coefficient of a term renders at least one of these three statements false. 
Wavefunction collapse is ``brutal'' in this regard, 
causing the coefficient of one term to vanish while forcing the coefficients of the remaining terms 
to increase in magnitude to preserve a unit norm for the wavefunction.

The next omitted assumption is unitarity, which we denote by (U). 
Reference \refcite{delRio} noticed this result.
If a non-unitary version of quantum mechanics is used, 
then again, one or more of the four ``If ..., then ...'' statements is no longer valid. 
Table 4 of reference \refcite{FRGE} provides a list of 10 interpretations/modifications 
of quantum mechanics, 
each of which violates at least one of the assumptions (Q), (C) and (S) used in the Frauchiger-Renner ``proof.''
However, most of these theories are non-unitary; the fact they violate (Q), (C) or (S) is secondary because
at least one the four ``If ..., then ...'' statements is not true. 
This second assumption, that of unitarity, is not fatal: 
one can simply add it as an additional assumption. 
Assumption (U) already includes the first omitted assumption concerning wavefunction collapse 
because wavefunction collapse violates unitarity. 

The third omitted assumption, which we label as (L), 
is that the generation of the contradiction assumes that ``standard classical'' logic 
can be applied to statements about quantum mechanical measurements. 
Indeed, the transitive property of logic needs to be used three times to generate the contradiction. 
R. Renner admits that he left assumption (L) out.\cite{RennerEmailA}
Although one can postulate that ``standard classical'' logical applies to quantum mechanics,\cite{RennerEmailB}
it is something that can be checked, 
and when one analyzes whether logical transitivity is obeyed in unitary quantum mechanics,
one finds that, in general, it is not. 
Hence, the derivation of the contradictory statement by Frauchiger and Renner is flawed. 
We also demonstrate that certain pairs of premises among the four ``If ..., then ...'' statements are incompatible, 
meaning that the ``If ..., then ...'' statements cannot all be logically used at once.
In short, the two authors are not showing that quantum mechanics cannot be applied to macroscopic systems;  
they are actually showing that ``classical'' logic can be violated in quantum mechanics. 

Intuitively, the reason for this is that the premises of the four ``If ..., then ...'' statements can interfere with each other. 
In our work on ``quantum logic'' in this article, we provide three explanations of why logical transitivity is violated, 
one of which is physically similar to the two-slit interference effect:
It is as though one of the ``If ..., then ..." statements requires ``the wavefunction to go through both slits'' to be true, 
while the other ``If ..., then ..." statement requires ``the wavefunction to go through only one of the slits''. 
In addition to showing that transitivity is violated in a direct calculation and presenting three ways 
to understand its violation, 
we provide below the proper way of combining two ``If ..., then ...'' measurement statements 
and a formula for quantifying its violation when the naive use of logical transitivity is applied; 
from this perspective, 
Frauchiger and Renner used the wrong method of combining ``If ..., then ...'' statements about measurements.  

A fourth omitted assumption is the following: 
The starting point of the Frauchiger-Renner gedanken experiment is the selection of a particular run: 
Of the four possible measurement outcomes that the two Wigner agents can obtain, 
Frauchiger and Renner run the experiment until only the `OK'--`$\overline{\mathrm{OK}}$' outcome occurs. 
The two authors implicitly assumed that the four statements remain valid for a particular run.\footnote{I initially overlooked this issue too. 
In the email correspondences between Renato Renner and myself 
from 3/11/21 to 3/21/22 (before submission of this work to Nature Communications) 
and from 4/15/22 to 5/21/22 (post submission), 
we both thought that statements 2 - 4 were valid for any particular run.}
However, it turns out that this assumption is also false. 
For the particular run selected, 
two of the ``If ..., then ...'' statements no longer remain true. 
The third one (Statement 4) is only ``technically'' true because its premise is always false,   
and this was also noted by D. Lazarovici and M. Hubert in Bohmian mechanics.\cite{LazaroviciHubert}
What happens is that when Agent W obtains a measurement of  `OK', 
it is impossible for Agent $\bar{\mathrm{F}}$ to obtain the outcome needed for the premise of Statement 4. 
Given that wavefunction collapse ruins the validity of at least one ``If ..., then ...'' statement, 
and that selecting a particular run is similar to wavefunction collapse, 
it is not surprising that this step of the Frauchiger-Renner gedanken experiment 
renders two of the ``If ..., then ...'' statements false. 
Instead of performing the gedanken experiment repeatedly until the two Wigner agents obtain the `OK'--`$\overline{\mathrm{OK}}$' outcome, 
one can phrase the first step of the gedanken experiment as a logic statement: 
``If Agent W and Agent $\bar{\mathrm{W}}$ obtain measurements of `OK' and `$\overline{\mathrm{OK}}$' respectively, then ...''. 
When formulated this way, assumption four is no longer needed, 
and the flaws in the Frauchiger-Renner analysis raised in this paragraph become violations of assumption (L). 

Frauchiger and Renner concluded from the contradictory statement that quantum mechanics 
cannot be applied to experimentalists and their equipment because 
the third and fourth steps in the gedanken experiment involved Wigner/friend measurements 
on these macroscopic systems. 
However, it could be that quantum mechanics {\it does} govern these macroscopic systems,  
but it is {\it impossible} to perform such measurements. 
Indeed, the Wigner/friend measurements used in the Frauchiger and Renner publication 
are of a very unusual nature, conducted on generalized Schr\"odinger cat states. 
Given the choice between whether (i) quantum mechanics can govern macroscopic systems 
and (ii) measurements can be performed on macroscopic  Schr\"odinger cat states, 
many physicists would assume that (ii) is impossible rather than (i). 

So the fifth omitted assumption is that the third and fourth measurements of the Gedanken experiment in reference \refcite{FRGE} 
are possible. 
However, in Section \ref{WignerMeasurement}, 
we point out that an overlooked agreement between a Wigner agent and the agent's friend 
needs to be established prior to the start of the Frauchiger-Renner gedanken experiment. 
This agreement is of interest because it puts restrictions on the nature of Wigner/friend experiments. 
Indeed, the restrictions are so stringent as to rule out Wigner measurements 
of the type used in the Frauchiger-Renner gedanken experiment on macroscopic entities. 
We consider this to be one of the important results of our work. 
It is interesting that Heisenberg's uncertainty principle plays a role in this.  
See Section \ref{WignerMeasurement}.
This restriction already casts doubt as to whether the Frauchiger-Renner gedanken experiment 
indicates that quantum mechanics cannot be extrapolated to complex systems. 

It is possible, however, to modify the Wigner/friend measurements so that the Heisenberg uncertainty principle is not violated.
When this is done, 
the third and fourth measurement only involve certain microscopic entities of the quantum constituents of the laboratory equipment of friend agents. 
The third and fourth measurements 
are then rendered ``ordinary'' (such as measurements on a spin using a particular axis of quantization). 
If assumption (L) were valid, one would then conclude the nonsensical result that quantum mechanics is inconsistent at a microscopic level. 
By modifying the Wigner/friend measurements as described below, 
it becomes possible to perform the Frauchiger-Renner experiment in a real laboratory setting, 
a result that should be of interest to experimentalists working on the foundations of quantum mechanics. 

In short, there a several flaws in the analysis of reference \refcite{FRGE}. 
There are also some issues with the assumptions (S) and (C) that are discussed in Appendix A. 
The problem is that they involve classical concepts. 


We use the Frauchiger-Renner gedanken experiment as a laboratory 
to explore a number of topics in quantum mechanics including 
wavefunction logic, Wigner/friend experiments, 
and the deduction of mathematical statements from knowledge of a wavefunction 
and obtain a number of interesting results. 

The validity of the Frauchiger-Renner gedanken experiment 
has been challenged\cite{LazaroviciHubert,SudberyA,SudberyB,Yang,LoeligerVontobel,MatzkinSokolovskiA,MatzkinSokolovskiB,WaaijerNeerven,Elouard}
in two directions: whether the argument itself is incorrect 
or whether there are hidden assumptions in the argument. 
However, no publication has pointed out that the fundamental reason why the analysis in the 2008 Nature Communications publication is wrong 
is that classical logic cannot be applied to Agent measurement statement.

In unitary quantum mechanics,\cite{SSMeasurement} measurement does not involve wavefunction collapse,\cite{EverettThesisA,EverettThesisB} 
the probability of an outcome is related to the absolute square of the wavefunction, 
and linearity and unitarity are strictly maintained even during a measuring process. 
There is a single universal wavefunction and one does not assign worlds to certain linear superpositions of this wavefunction. 
Instead, in unitary quantum mechanics, 
the quantum mechanical interpretation of a situation is obtained by examining the wavefunction itself;  
sections \ref{TheWavefunctions}, \ref{TheArgument} and \ref{WavefunctionLogic} illustrate this. 

Hence, unitary quantum mechanics is ``standard'' quantum mechanics without wavefunction collapse;
however, unitarity and the absence of wavefunction collapse lead to some consequences 
for quantum measurement that many physicists may not be unaccustomed to, 
which are embodied in the following: 
The basic {\bf Measurement Rule} is:\cite{SSMeasurement} 
{\it If wavefunction collapse is not needed ``to explain'' an experimental result, 
then a single measuring event suffices to determine the state with certainty;  
if this is not the case, then the uncertainty of the quantum state is transferred 
to the measuring agent, multiple measurements are needed to determine the state, 
and an output reading indicating that the state is $S$ does not mean that the wavefunction is $S$.}

In unitary quantum mechanics, 
one knows the {\it form} of the wavefunction 
at each stage of the Frauchiger-Renner gedanken experiment. 
From the wavefunction, 
it is quite easy to derive the four "If ... then ..." statements in reference \refcite{FRGE} 
This makes the analysis quite simple,  
and the violation of classical logic manifestly evident. 

All the measurements in the Frauchiger-Renner gedanken experiment involve binary outcomes. 
Let us illustrate the Measurement Rule in unitary quantum mechanics for this case. 
Let $\uspin$ and $\dspin$ indicate the two outcomes for a measurement,
and think of them as the up and down spin of a spin-$\oneHalf$ object. 
Let A be an agent who is about to measure the spin. 
The word agent included an experimentalist and her equipment. 
Let $\Psi^A$ be the agent's wavefunction before the measurement takes place. 
Then there are two generic cases: 
(i) The initial state is not a superposition; it is either $\uspin$ or $\dspin$ (up to an overall phase), 
but it is unknown which of these two possibilities is happening.
(ii) The initial state $S_0$ of the object to be measured is a superposition of up and down spin, that is, 
it is of the form $S_0 = a_{\uparrow} \uspin + a_{\downarrow} \dspin$ 
(with $|a_{\uparrow}|^2 +|a_{\downarrow}|^2 = 1$). 
In case (i), the schematic description of the process is 
\begin{equation}
\begin{aligned}
     \Psi^A \uspin \to \Psi_{\uparrow}^A  \textrm{ ,   if }    S_0 =  \uspin&  \\ 
                                                                                                         & \ \ \ \ \ \textrm{case (i) } \\
     \Psi^A \dspin \to \Psi_{\downarrow}^A  \textrm{ ,  if } S_0 =  \dspin&   
\ .
\label{DeterminedMeasurement} 
\end{aligned}
\end{equation}
Here, $\Psi_{\uparrow}^A$ and $\Psi_{\downarrow}^A$ are two different wavefunctions 
involving the quantum constituents of $A$ and the spin-$\oneHalf$ object. 
In the schematic equations, the left-hand side (respectively, right-hand side) 
is the wavefunction before (respectively, after) the measurement is made. 
Case (i) corresponds to the situation when wavefunction collapse is not needed to explain the experimental result: 
If the spin is up, then it is measured to be up and no wavefunction collapse is needed; 
ditto for the situation when the spin is down.
In case (ii), the schematic description in unitary quantum mechanics is
\begin{equation}
     \Psi^A (a_{\uparrow} \uspin + a_{\downarrow} \dspin) \to  a_{\uparrow}  \Psi_{\uparrow}^A + a_{\downarrow} \Psi_{\downarrow}^A 
 \ , \ \ \textrm{case (ii)} \ 
\ .
\label{UncertainMeasurement} 
\end{equation}
Equation (\ref{UncertainMeasurement}) follows from Eq.(\ref{DeterminedMeasurement}) and quantum-mechanical linearity, 
and linearity is a consequence of unitarity. 
If Agent A is a machine with no thinking capability, 
then $\Psi_{\uparrow}^A$ in Eq.(\ref{UncertainMeasurement}) is the same as  $\Psi_{\uparrow}^A$ in Eq.(\ref{DeterminedMeasurement}) 
and its quantum constituents involve the coding of an up-spin output; likewise for $\Psi_{\downarrow}^A$.
If Agent A involves a human or entities with reasoning ability, 
and this is the case for the agents in the Frauchiger-Renner gedanken experiment, 
then the relation between the wavefunctions in Eqs.(\ref{UncertainMeasurement}) and (\ref{DeterminedMeasurement}) 
depends on what Agent A knows about the initial state. 
If Agent A knows nothing, then the situation is the same as that of the pure machine case: 
the wavefunction components, $\Psi_{\uparrow}^A$ and $\Psi_{\downarrow}^A$, 
in equations (\ref{DeterminedMeasurement}) and (\ref{UncertainMeasurement}) are equal. 
If Agent A knows that the initial situation is (i), then $\Psi_{\uparrow}^A$ contains a configuration of the human's quantum constituents 
that embody the thought ``I measured the spin to be up and so I know the initial wavefunction must have been $S_0 = \uspin$.''
A similar thought holds with ``up'' replaced by ``down'' is contained in $\Psi_{\downarrow}^A$. 
If Agent A knows that the initial situation is (ii) but is using unitary quantum mechanics, then 
$\Psi_{\uparrow}^A$ contains a configuration of the human's quantum constituents 
that embodies the thought ``I measured the spin to be up but I know that the current wavefunction 
must contain another component $\Psi_{\downarrow}^A$ in a linear superposition 
even though I cannot be directly aware of its existence.''
A similar statement arises for $\Psi_{\downarrow}^A$. 
In case (ii), even though $\Psi_{\uparrow}^A$ involves the observation of an output indicating up spin, 
Agent A cannot conclude that the spin is or was up. 
Indeed, this is correct because the spin was initially $a_{\uparrow} \uspin + a_{\downarrow} \dspin$ (and not just $\uspin$)
and it is very unlikely to end up being proportional to $\uspin$ at any point during a physical measuring process. 
Furthermore, it takes multiple measurements beginning with the same $S_0 = a_{\uparrow} \uspin + a_{\downarrow} \dspin$ 
to determine information about the coefficients $ a_{\uparrow}$ and $ a_{\downarrow}$. 

The agents in the Frauchiger-Renner gedanken experiment are not only informed of the initial wavefunction 
but also of the entire series of measurement steps. 
In relation to what was discussed in the previous paragraph, the analog is as follows:
If Agent A was informed that $S_0$ was $a_{\uparrow} \uspin + a_{\downarrow} \dspin$, 
then, in this case, 
$\Psi_{\uparrow}^A$ contains a configuration of the human's quantum constituents 
that embodies the thought ``I measured the spin to be up but I know that the current wavefunction 
must contain another component $\Psi_{\downarrow}^A$ in a linear superposition 
even though I cannot be directly aware of its existence {\it and} that the form of this superposition 
is $a_{\uparrow}  \Psi_{\uparrow}^A + a_{\downarrow} \Psi_{\downarrow}^A$, 
where $a_{\uparrow}$ and $a_{\downarrow}$ are the same coefficients as in $S_0$.''
Furthermore, another Agent B, having been given all the information about the initial state and experimental procedure, 
can also conclude that the form of the wavefunction after the measurement 
is $a_{\uparrow}  \tilde{\Psi}_{\uparrow}^A + a_{\downarrow} \tilde{\Psi}_{\downarrow}^A$, 
where $\tilde{\Psi}_{\uparrow}^A$ and $\tilde{\Psi}_{\downarrow}^A$ are some wavefunctions 
that Agent B does not know in detail but which incorporate the same measurement statements 
as in ${\Psi}_{\uparrow}^A$ and ${\Psi}_{\downarrow}^A$: 
``Agent A measured the spin to be up (or down) but Agent A knows that the current wavefunction must contain another ... ''.
Agent B can be one of the agents in the Frauchiger-Renner gedanken experiment 
or even an ``outsider'', that is, someone who is not directly involved. 
Reference \refcite{FRGE} thought that an assumption (Assumption (C)) was needed 
for agents such as Agent B to make such deductions. 
In unitary quantum mechanics, Assumption (C) is not needed 
because the ``If ..., then '' statements can be derived without the need of this assumption. 
See Appendix A. 

Our notation differs somewhat from that of Frauchiger and Renner:\cite{FRGE} 
The use of ``bars'' over objects is unchanged.
However, we use ``bra's'' and ``ket's'' to denote discrete states, 
which happen to come in pairs for the Frauchiger-Renner gedanken experiment;
so, we denote them with up and down spins ($\uspin$ and $\dspin$)
to take advantage of the isomorphism with spin-$\oneHalf$ objects. 
For states involving many degrees of freedom, 
we denote the wavefunction using the symbol $\Psi$.
A subscript $M$ on a state indicates that it has been ``measured'' 
but  $M$ can also stand for ``message'' because these states also embody statements 
such as the ones in quotes in the two previous paragraphs. 
The states $\uspin_S$ and $\dspin_S$ in reference \refcite{FRGE} are simply denoted by $\uspin$ and $\dspin$ in our paper.
Our states $\uspinbar$ and $\dspinbar$ correspond to $\ket{\textrm{heads}}_R$ and $\ket{\textrm{tails}}_R$ in ref.\refcite{FRGE}.
A discrete state associated with ``ok'' (respectively, ``fail'')  in ref.\refcite{FRGE} 
is represented by a `$-$' (respectively, a `$+$') in our work
(except our unbarred $\ket{-}_M$ corresponds to $ -\ket{\textrm{ok}}_{\textrm{L}}$, that is, it differs by a minus sign). 
The measurement times $t_j$ in our paper are written as n:0j in reference \refcite{FRGE}.

\section{Wavefunction Representations of the Extended Wigner/Friend Experiment in Unitary Quantum Mechanics}
\label{TheWavefunctions}

The initial state $\Psi_0$ of the Frauchiger-Renner gedanken experiment can be taken to be\footnote{In reference \refcite{FRGE}, 
the initial state is created differently: 
A quantum qubit of the form $\frac{1}{\sqrt{2}} \uspinbar +  \sqrt{\frac{2}{3}} \dspinbar$ is generated.
Agent $\bar{\mathrm{F}}$ measures this barred spin state. 
If the barred spin is up, then she sends the state $\dspin$ to Agent F. 
If it is down, then she sends  the state $(\uspin + \dspin)/\sqrt{2}$ to Agent F.
This procedure assumes that $\bar{\mathrm{F}}$ is able to manipulate states easily. 
In particular, the overall phase of a wavefunction is not an observable and cannot be controlled,
and so she cannot guarantee that $(\uspin + \dspin)/\sqrt{2}$ as opposed to $e^{i \phi}(\uspin + \dspin)/\sqrt{2}$ is sent. 
The resulting initial wavefunction would become 
$(\uspinbar \dspin + e^{i \phi} \dspinbar (\uspin + \dspin))/\sqrt{3}$.
Put differently, Agent $\bar{\mathrm{F}}$ cannot control the relative phase of the two terms 
when the procedure in reference \refcite{FRGE} is used. 
Statement 2 below is not true unless $\phi = 0$. 
Hence, it is better to begin with Eq.(\ref{InitialState}).
}
\begin{equation}
     \Psi_0 =    \frac{\Psi^W \psibar{W} \Psi^F \psibar{F} }{\sqrt{3}} 
        (\uspinbar \dspin + \dspinbar \uspin + \dspinbar \dspin)
\ .
\label{InitialState} 
\end{equation}
This state involves two qubits, which we take to be two spin-$\oneHalf$ objects: 
an ``unbarred'' spin, for which the basis is $\uspin$ and $\dspin$, 
and a ``barred'' spin, for which the basis is $\uspinbar$ and $\dspinbar$. 
In both cases, the axis of quantization for the spin is taken to be in the positive $z$-direction.

The Gedanken experiment proceeds in four main measurement steps: 

In the first step, Agent $\bar{\mathrm{F}}$ measures the spin of the barred spin-$\oneHalf$ object in the $z$-direction at time $t_1$. 
Using Eq.(\ref{UncertainMeasurement}), this causes $\psibar{F} \uspinbar$ and $\psibar{F} \dspinbar$ to be replaced by 
$\psibar{F}_{\uparrow}$ and $\psibar{F}_{\downarrow}$ respectively.
The wavefunction then becomes 
\begin{equation}
     \Psi_1 =   \frac{\Psi^W \psibar{W} \Psi^F }{\sqrt{3}} 
          (\uspinbar_M \dspin + \dspinbar_M \uspin + \dspinbar_M \dspin)
\ .
\label{StateTime1} 
\end{equation}
In Eq.(\ref{StateTime1}), we have replaced $\psibar{F}_{\uparrow}$ and $\psibar{F}_{\downarrow}$ 
by $\uspinbar_M$ and $\dspinbar_M$ respectively. 
They can be considered to make up a discrete two-state system with a message associated with each. 
The justification for this is given in Section \ref{WignerMeasurement}, which discusses some aspects of Wigner/friend experiments 
in the context of the Frauchiger-Renner gedanken experiment. 
This replacement does not affect any of the conclusions obtained in our paper 
concerning classical logic in quantum mechanics. 
A reader who does not want to use this simplification can replace $\uspinbar_M$ and $\dspinbar_M$  
with $\psibar{F}_{\uparrow}$ and $\psibar{F}_{\downarrow}$ in the equations below.

In the second step, Agent F measures the spin of the unbarred spin-$\oneHalf$ object at time $t_2$ 
in a way that is analogous as to what Agent  $\bar{\mathrm{F}}$ did for barred spin at time $t_1$, 
and the wavefunction becomes
\begin{equation}
     \Psi_2 =   \frac{\Psi^W  \psibar{W} }{\sqrt{3}} 
         (\uspinbar_M \dspin_M + \dspinbar_M \uspin_M + \dspinbar_M \dspin_M)
\ .
\label{StateTime2} 
\end{equation}

The third step involves a Wigner measurement by Wigner Agent $\bar{\mathrm{W}}$ 
of `measured' barred spin in the $x$-direction.
Here, we are thinking of $\uspinbar_M$ and  $\dspinbar_M$ 
as the up and down $z$-components of a spin-$\oneHalf$ system.
More precisely, the measurement is performed on the basis 
$\ket{\overline{+}}_M = (\uspinbar_M + \dspinbar_M)/\sqrt{2}$ (spin in the positive $x$-direction) 
and 
$\ket{\overline{-}}_M = (\uspinbar_M - \dspinbar_M)/\sqrt{2}$ (spin in the negative $x$-direction).
Expressing $\uspinbar_M$ as $ (\ket{\overline{+}}_M + \ket{\overline{-}}_M)/\sqrt{2}$ 
and $\dspinbar_M$ as $ (\ket{\overline{+}}_M - \ket{\overline{-}}_M)/\sqrt{2}$, 
one arrives at 
\begin{equation}
     \Psi_3 = \frac{\Psi^W}{\sqrt{6}} 
          {\big( (\psibar{W}_+ + \psibar{W}_-) \dspin_M + (\psibar{W}_+ - \psibar{W}_-) \uspin_M + (\psibar{W}_+ - \psibar{W}_-)  \dspin_M \big)}
\ , 
\label{StateTime3} 
\end{equation}
as the wavefunction after Agent $\bar{\mathrm{W}}$ makes the measurement at time $t_3$.

Finally, Agent W performs a similar measurement (that is, in the $x$-direction for `measured' unbarred spin) 
as Agent $\bar{\mathrm{W}}$ but on the $\uspin_M$ and $\dspin_M$ 
with the result 
\begin{equation}
\begin{aligned}
     \Psi_4 = \frac{1}{\sqrt{12}}  \big(   (\psibar{W}_+ + \psibar{W}_-) & ( \Psi^W_+ -  \Psi^W_-) 
          +   (\psibar{W}_+ - \psibar{W}_-)  ( \Psi^W_+ +  \Psi^W_-)    \\ 
        + & (\psibar{W}_+ - \psibar{W}_-)  ( \Psi^W_+ -  \Psi^W_-) 
   \big)
\ , 
\label{StateTime4} 
\end{aligned}
\end{equation}
at the final time $t_f = t_4$. 
The first term in Eq.(\ref{StateTime4}) comes from the first term $\uspinbar \dspin$ in Eq.(\ref{InitialState}), 
the second term from the second term $\dspinbar \uspin $ in Eq.(\ref{InitialState})
and the third from the third one $\dspinbar \dspin$. 
Some terms in Eq.(\ref{StateTime4}) cancel among themselves 
(which can be considered a quantum interference effect) to give 
\begin{equation}
     \Psi_4 = \frac{1}{\sqrt{12}}  \big(  
       3 \psibar{W}_+ \Psi^W_+ - \psibar{W}_+ \Psi^W_- - \psibar{W}_- \Psi^W_+ - \psibar{W}_- \Psi^W_-
   \big)
\ .
\label{StateTimeFinal} 
\end{equation}
Equations \ref{StateTime2} -  \ref{StateTimeFinal} are consistent with the results obtained
in references \refcite{LazaroviciHubert} - \refcite{SudberyB} and \refcite{MatzkinSokolovskiA} - \refcite{WaaijerNeerven}
after one takes into account notational differences and/or the use of state-preserving\footnote{This is a particular form of measurement 
in which the measured state is preserved:  
\newline
$
     \Psi^A (a_{\uparrow} \uspin + a_{\downarrow} \dspin) \to 
    a_{\uparrow}  \Psi_{\uparrow}^A  \uspin+ a_{\downarrow} \Psi_{\downarrow}^A \dspin
$, 
where $A$ is a ``friend agent'', that is, $F$ or $\bar{F}$, or
where $A$ is a ``Wigner agent'', that is, $W$ or $\bar{W}$ and 
$\uparrow$ is replaced by ${+}$ and $\downarrow$ is replaced by ${-}$.
Note that the difference between this and Eq.(\ref{UncertainMeasurement}) is that 
the states $\uspin$ and $\dspin$ remain unchanged after the measurement. 
One might call this type of measurement an observation: 
An agent observes the state but leaves it intact.
The ``measurement information statement'' is still in $\Psi_{\uparrow}^A$ and $ \Psi_{\downarrow}^A$. }
measurements performed by the agents. 

There is an additional step in which Agent W meets with Agent $\bar{\mathrm{W}}$ 
and they provide each other with their measurement results. 
This affects both of their wavefunctions yielding: 
\begin{equation}
     \Psi_5 = \frac{1}{\sqrt{12}}  \big(  
       3 \psibar{W}_{++} \Psi^W_{++} - \psibar{W}_{+-} \Psi^W_{+-} - \psibar{W}_{-+} \Psi^W_{-+} - \psibar{W}_{--} \Psi^W_{--}
   \big)
\ .
\label{StateTimeFinal2} 
\end{equation}
A subscript ${xy}$ on a $\Psi$ encodes the statement ``Agent $\bar{\mathrm{W}}$ measured $x$ at time $t_3$ and Agent W measured $y$ at time $t_f$.''
For example, $\psibar{W}_{-+}$ involves ``I, Agent $\bar{\mathrm{W}}$, measured `$-$' at time $t_3$ 
and subsequently met with Agent W and `learned' that Agent W had measured `$+$'  at time $t_f$.

\section{The Frauchiger-Renner Argument}
\label{TheArgument}

The wavefunctions in Section \ref{TheWavefunctions} encode measurement statements about themselves: 

$\uspinbar_M$ with the statement ``Agent $\bar{\mathrm{F}}$ measured the barred spin to be up (that is, $\uspinbar$) at time $t_1$'',  

$\dspinbar_M$ with ``Agent $\bar{\mathrm{F}}$ measured the barred spin to be down at time $t_1$'',  

$\uspin_M$ with ``Agent F measure the unbarred spin to be up at time $t_2$'', 

$\dspin_M$ with ``Agent F measure the unbarred spin to be down at time $t_2$'', 

$\psibar{W}_+$ with ``Agent  $\bar{\mathrm{W}}$ obtained a measurement of `$\overline{+}$' a time $t_3$'',

$\psibar{W}_-$ with ``Agent  $\bar{\mathrm{W}}$ obtained a measurement of `$\overline{-}$' a time $t_3$'',

$\Psi^W_+$ with ``Agent W obtained a measurement of `$+$' a time $t_4$'', and 

$\Psi^W_-$ with ``Agent W obtained a measurement of `$-$' a time $t_4$.''

The Frauchiger-Renner argument that quantum mechanics cannot consistently describe itself is based 
on four statements that can be derived from the wavefunction results in Section  \ref{TheWavefunctions}. 

Statement 1: When the experiment is carried out multiple times, 
there is eventually a run in which Agent W measures `$-$' and when he encounters Agent $\bar{\mathrm{W}}$ 
the latter informs the former that he has measured `$\overline{-}$'. 
For this particular run, Agent W can say ``If I (Agent W) measured `$-$' at time $t_4$, then 
Agent $\bar{\mathrm{W}}$ measured `$\overline{-}$' at time $t_3$.''

However, it is convenient to consider an alternative form of Statement 1:\\
Statement 1$'$: ``If, at time $t_4$, agents W and $\bar{\mathrm{W}}$ respectively obtain 
`$-$' {\it and} `$\overline{-}$', then $\bar{\mathrm{W}}$ obtained `$\overline{-}$'.'' \\
This modified statement is trivially true. 
One also needs to note that the probability of them both measuring a `$-$' result is non-zero. 
The advantage of using this form is that one does not have to repeatedly run the experiment 
until the ``minus-minus'' outcome happens. 

Statement 2: If Agent $\bar{\mathrm{W}}$ measured `$\overline{-}$' at time $t_3$, then Agent F measured 
the unbarred spin to be up at time $t_2$. 

Statement 3: If Agent F measured the unbarred spin to be up at time $t_2$,  
then Agent $\bar{\mathrm{F}}$ measured the barred spin to be down at time $t_1$.

Statement 4: If Agent $\bar{\mathrm{F}}$ measured the barred spin to be down at time $t_1$, then 
Agent W will measure `$+$' at time $t_4$. 

Using the transitive property of logic and 
combining Statements 1 to 4 in order for the run in which both Wigner agents get ``minus''  
produces a contradiction: 
The ``If ... then ...'' logical statement begins with the premise ``Agent W measured `$-$' at time $t_4$'' 
and ends with the conclusion ``Agent W will measure `$+$' at time $t_4$,'' 
which we call the {\it Contradictory Statement}.\footnote{When using Statement 1$'$, 
the Contradictory Statement becomes ``If, at time $t_4$, agents W and $\bar{\mathrm{W}}$ respectively measure 
`$-$' {\it and} `$\overline{-}$', then agent W can deduce that he will measure `$+$'.}
Notice that putting these statements together does not correspond to the order in which measurements take place. 
However, this does not matter in unitary quantum mechanics. 
Agents W and $\bar{\mathrm{W}}$ can deduce the four statements all at time $t_4$.
In fact, anybody who knows the initial state and the experimental procedure can deduce the four statements 
including those ``If ..., then ...'' statements whose premise involve a time later than its conclusion.  
See Appendix A. 
Based on the Contradictory Statement, Frauchiger and Renner conclude that quantum mechanics cannot consistently describe itself. 

Statement 1 (or Statement 1$'$) follows from Eq.(\ref{StateTimeFinal}): there is a term $-\frac{1}{\sqrt{12}} \psibar{W}_- \Psi^W_-$ in the final wavefunction 
meaning that $\frac{1}{12}$th of the time Agent W measures `$-$' 
and Agent $\bar{\mathrm{W}}$  measures `$\overline{-}$'. 

Statement 2 follows from Eq.(\ref{StateTime3}): If one looks at the term proportional to $\psibar{W}_-$,  
then the first and third terms involving $\dspin_M$ cancel leaving only the term $-\frac{1}{\sqrt{6}} \psibar{W}_- \uspin_M$. 

Statement 3 follows from Eq.(\ref{StateTime2}): The term proportional to $\uspin_M$ only involves $\dspinbar_M$.
There is no $\uspin_M \uspinbar_M$ term in the wavefunction.

Statement 4 follows from Eqs.(\ref{StateTime1}) and (\ref{StateTime4}): 
If Agent $\bar{\mathrm{F}}$ measured the barred spin to be down (that is, $\dspinbar$) at time $t_1$, then 
the relevant part of the wavefunction consists of the second and third terms in Eq.(\ref{StateTime1}). 
These two terms evolve to the second and third terms in Eq.(\ref{StateTime4}) 
but the term proportional to $ \Psi^W_-$ cancels among them. 

It is also possible to {\it experimentally} verfiy each of the above four statements by introducing two additional agents.
See the discussion at the end of the next section. 

If wavefunction collapse occurs, then the wavefunctions are not given as in Eqs.(\ref{StateTime1}) - (\ref{StateTime4}) 
but depend on the outcomes of the agents' measurements. 
For example, if in the first measurement Agent $\bar{\mathrm{F}}$ measures the barred spin to be up, 
then the wavefunction collapses to the first term in Eq.(\ref{StateTime1}) at time $t_1$. 
Ignoring for the moment the effects of the measurements of the other agents, 
Statement 2 is no longer true because its validity depended on a cancellation of the first and third terms, 
but the latter is now missing. 
Statements 3 and 4 are technically true only because their premises are false. 
If Agent $\bar{\mathrm{F}}$ measures the barred spin to be down, 
then the wavefunction collapses to the second and third terms in  in Eq.(\ref{StateTime1}) at time $t_1$, 
again rending Statement 2 false; Statements 3 and 4 remain true. 
If Agent $\bar{\mathrm{F}}$ measures an the barred spin to be down at time $t_1$
and Agent F measures the unbarred spin to be down at time $t_2$, 
then only the last term remains in the wavefunction in Eq.(\ref{StateTime2}). 
In this case, 
Statements 2 and 4 are no longer true 
and Statement 3 is ``logically true'' only because its premise is false. 
The above is illustrative of what is true in general: 
wavefunction collapse renders one or more of statements 2 through 4 false.

\section{Quantum Logic and Mathematical Wavefunction Statements}
\label{WavefunctionLogic}

In this section, we discuss the logical mathematical statements encoded in wavefunctions. 
Although individual mathematical statements have a physical analog in terms of experimental measurements, 
sets of mathematical statements do not necessarily have a physical analog in terms of a sequence of steps in an experiment, 
as will become clear below. 
The purely mathematical results presented in this section  
provide insights into the Frauchiger-Renner gedanken experiment
and the issue of violation of the transitive property of logic in quantum mechanics in general. 
It is worth noting that experimentalists can make measurements to separately verify each of the mathematical statements about wavefunctions; 
this is shown near the end of this section; see the paragraph staring with Eq.(\ref{InitialStateVerificationzz}) and subsequent paragraphs. 

There are three basic rules: \\
{\bf (i)} When a wavefunction is written as a linear superposition of several orthogonal component wavefunctions, 
then the situation involves logical disjunction and the OR symbol.\footnote{{It is not possible 
to have a logical conjuction (AND symbol) between two (or more) orthogonal components 
of a wavefunction. 
In Eq.(\ref{InitialStateLogic}), 
the logical statement $({\uspinbar}_z {\dspin}_z$ AND ${\dspinbar}_z {\uspin}_z)$ makes no sense: 
It is impossible for the barred spin to be up (and unbarred spin down) and for the  the barred spin to be down (and unbarred spin up). 
The fact that linear superpositions of orthogonal wavefunctions have this property is the reason 
why Schr\"odinger cats are non-problematic in unitary quantum mechanics.\cite{SSMeasurement}}} \\
{\bf (ii)} When a wavefunction involves a product of several component wavefunctions, 
then the situation involves logical conjunction and the AND symbol. \\
{\bf (iii)} The probability that a (normalized) state occurs is given by 
the absolute square of its coefficient in the wavefunction (the Born rule). 

Let us illustrate these rules using the spin part of the wavefunction at time $t_0$ in Eq.(\ref{InitialState}):
\begin{equation}
\Psi =  \frac{1}{\sqrt{3}}{\uspinbar}_z {\dspin}_z +  \frac{1}{\sqrt{3}}{\dspinbar}_z {\uspin}_z + \frac{1}{\sqrt{3}}{\dspinbar}_z {\dspin}_z 
\ .
\label{InitialStateLogic}
\end{equation}
Rule (i) tells us that either the situation is ${\uspinbar}_z {\dspin}_z$ OR ${\dspinbar}_z {\uspin}_z$ OR $ {\dspinbar}_z {\dspin}_z$.
This particular case not only involves logical disjunction but mutual exclusivity.$^e$ 
If the wavefunction were only  ${\uspinbar}_z {\dspin}_z$, then Rule (ii) would tell us 
that the barred spin is up AND the unbarred spin is down. 
As a more complicated example, 
combining rules (i) and (ii), one arrives at the following mathematical statement from the wavefunction in Eq.(\ref{InitialStateLogic}):
\begin{equation}
\begin{aligned}
\textrm{Either (barred spin is up AND the unbarred spin is down) OR} \\ 
\textrm{         (barred spin is down AND the unbarred spin is up) OR}  \\
\textrm{         (barred spin is down AND the unbarred spin is down)}
\ .
\label{InitialStateLogicStatement}
\end{aligned}
\end{equation}
One can also derive 
\begin{equation}
\begin{aligned}
\textrm{Mathematical Statement 3:}& \\
\textrm{If the unbarred spin is }&  {\uspin}_z \textrm{, then the barred spin is } {\dspinbar}_z
\ ,
\label{LogicStatement3}
\end{aligned}
\end{equation}
because only the middle term has the unbarred spin being up in Eq.(\ref{InitialStateLogic}).

Different logical statements can be derived by expanding the wavefunction in different ways.
For example, the second and third terms in Eq.(\ref{InitialStateLogic}) combine to give
\begin{equation}
\Psi = \frac{1}{\sqrt{3}}{\uspinbar}_z {\dspin}_z +  \sqrt{\frac{2}{3}}{\dspinbar}_z {\uspin}_x  
\ ,
\label{StateLogic4}
\end{equation}
where the $z$ and $x$ subscripts indicate the direction of spin quantization and
where ${\uspinbar}_x = ( {\uspinbar}_z + {\dspinbar}_z)/\sqrt{2}$.  
From the second term, one deduces 
\begin{equation}
\begin{aligned}
\textrm{Mathematical Statement 4:}& \\ 
\textrm{If the barred spin is }& {\dspinbar}_z \textrm{, then the unbarred spin is }  {\uspin}_x
\ .
\label{LogicStatement4}
\end{aligned}
\end{equation}
The violation of the transitive property of logic for mathematical statements derived from wavefunctions 
follows from Mathematical Statements 3 and 4: 
Let $A = ``$if the unbarred spin is ${\uspin}_z$'', 
$B = ``$the barred spin is $ {\dspinbar}_z$'', and $C = ``$the unbarred spin is ${\uspin}_x$''.
Putting these two statements (that is, $A \Rightarrow B$ and $B \Rightarrow C$) together 
using the transitive property of logic yields 
\begin{equation}
\textrm{If the unbarred spin is } {\uspin}_z  \textrm{, then the unbarred spin is } {\uspin}_x
\ ,
\label{RotationBy90}
\end{equation}
which is a contradiction. 
As we show below, 
the problem with transitivity for Mathematical Statements 3 and 4 in this paragraph is closely related to 
the problem with logic in combining Frauchiger-Renner's measurement Statements 3 and 4, 
but with $\ket{+}_M$ playing the role of ${\uspin}_x$.

Rule (iii) tells us that the probability of barred spin being up AND unbarred spin being down 
is $1/3$ because the coefficient of ${\uspinbar}_z {\dspin}_z$ is $1/\sqrt{3}$ in Eq.(\ref{InitialStateLogic}). 
Using Rule (iii), one can obtain mathematical wavefunction statements involving probabilities: 
\begin{equation}
\begin{aligned}
\textrm{Math}&\textrm{ematica Statement 4}'\textrm{:} \\ 
     &\textrm{If the barred spin is } {\dspinbar}_z \textrm{, then} \\
     &\textrm{\ \ \  there is a 50\% chance that unbarred spin is up } ({\uspin}_z), \textrm{ and } \\
     &\textrm{\ \ \  there is a 50\% chance that unbarred spin is down }  ({\dspin}_z).
\label{LogicStatement4p}
\end{aligned}
\end{equation}
This follows from the second and third terms in Eq.(\ref{InitialStateLogic}). 
Given that quantum mechanics is a theory of probability, 
it is natural for logical mathematical statements about wavefunction to involve probabilities. 
Now, when Mathematical Statements 3 and 4$'$ are combined using transitivity, 
they generate an invalid probabilistic statement:
\begin{equation}
\begin{aligned}
     &\textrm{If the unbarred spin is } {\uspin}_z \textrm{, then} \\
     &\textrm{\ \ \  there is a 50\% chance that unbarred spin is up } ({\uspin}_z), \textrm{ and } \\
     &\textrm{\ \ \  there is a 50\% chance that unbarred spin is down } ({\dspin}_z) 
\ ,
\label{RotationBy90p}
\end{aligned}
\end{equation}
which is consistent with the incorrect statement in Eq.(\ref{RotationBy90}) because a spin in the up $x$-direction 
has a 50\% chance of being up in the $z$-direction and
 a 50\% chance of being down in the $z$-direction.

By expressing the wavefunction in Eq.(\ref{InitialStateLogic})  
in an axis of spin quantization in the $z$ direction for the unbarred spin and in an $x$ direction axis for the barred spin, 
one can also obtain the mathematical statement about wavefunctions 
that is analogous to the Frauchiger-Renner measurement Statement 2:
\begin{equation}
\begin{aligned}
\textrm{Mathematical Statement 2:}& \\ 
\textrm{If the barred spin is }& {\dspinbar}_x \textrm{, then the unbarred spin is }  {\uspin}_z
\ .
\label{LogicStatement2}
\end{aligned}
\end{equation}
There is also the analog of Frauchiger-Renner Statement 1:
\begin{equation}
\begin{aligned}
\textrm{Mathematica} & \textrm{l Statement 1:} \\ 
\textrm{If unbarred} & \textrm{ and barred spins are respectively } {\dspin}_x  \textrm{ and } {\dspinbar}_x \ , \\
                                                             &\textrm{then the barred spin is } {\dspinbar}_x
\ ,
\label{LogicStatement1}
\end{aligned}
\end{equation}
which is obviously true. 

When all four mathematical statements (Eqs.(\ref{LogicStatement1}), (\ref{LogicStatement2}), (\ref{LogicStatement3}) and (\ref{LogicStatement4})) 
are combined using classical logic, one can then rotate the unbarred spin to flip it completely 
(rather than rotating it 90$^{\textrm{o}}$ as is done in Eq.(\ref{RotationBy90})). 
The chain of statements reads ``If unbarred and barred spins are respectively ${\dspin}_x$ and ${\dspinbar}_x$,  
then the barred spin is ${\dspinbar}_x$,''
``If the barred spin is ${\dspinbar}_x$, then the unbarred spin is ${\uspin}_z$,'' 
``If the unbarred spin is ${\uspin}_z$, then the barred spin is ${\dspinbar}_z$,''
and ``If the barred spin is ${\dspinbar}_z$, then the unbarred spin is ${\uspin}_x$."
If these could be combined using the transitive property of logic, then one would obtain the Contradictory Mathematical Statement 
about wavefunctions:
``If unbarred and barred spins are respectively ${\dspin}_x$ and ${\dspinbar}_x$, then the unbarred spin is  ${\uspin}_x$.''
The fact that there is a one-to-one correspondence between mathematical statements about the wavefunction 
in Eq.(\ref{InitialStateLogic}) and the Frauchiger-Renner measurement statements beginning with the wavefunction in Eq.(\ref{InitialState})
(whose spin component is identical)  
suggests that problem with the transitive property of logic for mathematical statements derived from wavefunctions 
is likely to arise in the Frauchiger-Renner gedanken experiment, 
but this remains to be shown, and we show this below. 

Summarizing, the transitive property of logic cannot always be used for mathematical statements derived from wavefunctions.
The question arises as to whether the above discussion can be ``translated'' into statements about measurements. 

Mathematical Statements 3 and 4 have physical equivalents involving measurement statements: 
\begin{equation}
\begin{aligned}
&\textrm{If the unbarred spin is measured to be } {\uspin}_z , \\
&\textrm{   then the barred spin will be measured to be } {\dspinbar}_z
\ ,
\label{LogicStatement3m}
\end{aligned}
\end{equation}
and
\begin{equation}
\begin{aligned}
&\textrm{If the unbarred spin is measured to be } {\dspinbar}_z , \\
&\textrm{   then the unbarred spin  will be measured to be }  {\uspin}_x
\ .
\label{LogicStatement4m}
\end{aligned}
\end{equation}
From the above, 
it might seem easy to create a physical experiment that generates a contradiction. 
This is not the case for two reasons: 
First, the measurements occur at different times. 
If the measurement of the unbarred spin in Eq.(\ref{LogicStatement3m}) happens at time $t_1$ 
and that of the barred spin in Eqs.(\ref{LogicStatement3m}) and (\ref{LogicStatement4m}) at time $t_2$,  
then the second unbarred spin measurement in Eq.(\ref{LogicStatement4m}) must necessarily happen after $t_2$ 
since the conclusion of the ``If ... then ...'' statement in Eq.(\ref{LogicStatement4m})) is in the future. 
Suppose it happens at time $t_3$.
Then combining Eqs.(\ref{LogicStatement3m}) and (\ref{LogicStatement4m}) using
the transitive property of logic gives
\begin{equation}
\begin{aligned}
&\textrm{If the unbarred spin is measured to be } {\uspin}_z  \textrm{ at time } t_1 ,\\
&\textrm{   then the unbarred spin  will be measured to be }  {\uspin}_x \textrm{ at time } t_3
\ ,
\label{RotationBy90m}
\end{aligned}
\end{equation}
which is not obviously a contradiction because the times are different. 
Second, measurements can affect wavefunctions. 
The measurement of the unbarred spin at time $t_1$ may disturb it, 
changing it from ${\uspin}_z$ to some other value in an unpredictable way.  
Hence, the subsequent measurement of the unbarred spin at time $t_3$ could have a random relation to its value at time $t_1$.

Statements 3 and 4 of the Frauchiger-Renner gedanken experiment 
only involve the second and third terms of Eq.(\ref{InitialState}). 
Let us focus on them and the measurements by agents F and Agent $\bar{\mathrm{F}}$: 
\begin{equation}
     \Psi_0 =    \frac{\Psi^F \psibar{F} }{\sqrt{2}} (\dspinbar \uspin + \dspinbar \dspin)
\ .
\label{InitialState23} 
\end{equation}
When Agent $\bar{\mathrm{F}}$ makes her measurement at $t_1$, 
the wavefunction becomes 
\begin{equation}
     \Psi_1 =   \frac{\Psi^F }{\sqrt{2}} 
          (\psibar{F}_{\downarrow} \uspin + \psibar{F}_{\downarrow} \dspin)
\ ,
\label{State23Time1} 
\end{equation}
and when Agent F makes her measurement at $t_2$,
it becomes
\begin{equation}
     \Psi_2 =   \frac{1}{\sqrt{2}} 
          (\psibar{F}_{\downarrow} \Psi^{F}_{\uparrow} + \psibar{F}_{\downarrow} \Psi^{F}_{\downarrow})
\ .
\label{State23Time2} 
\end{equation}
Now one has two valid statements, the original one of the Frauchiger-Renner gedanken experiment, which is,

Statement 3: If Agent F measured the unbarred spin to be up at time $t_2$,   
then Agent $\bar{\mathrm{F}}$ measured the barred spin to be down at time $t_1$, \\ 
and the analog of Mathematical Statement 4$'$, namely, 

Statement 4m: If Agent $\bar{\mathrm{F}}$ measured the barred spin to be down at time $t_1$,  
then Agent F will measure at time $t_2$ the unbarred spin to be up with a probability of 50\% 
and will measure it to be down with a probability of 50\%.

The conclusion of Statement 3 is the premise of Statement 4m. 
Hence, if {\it the transitive property of logic is valid for combining statements about measurements}, then 
Agent F can say, ``If I measure unbarred spin to be up at time $t_2$, 
then I can conclude that I am not guaranteed to measure the unbarred spin to be up at time $t_2$.'' 
Since this is a contradiction, the premise of the statement, namely, 
 {\it the transitive property of logic is valid for combining statements about measurements}, must be false. 
Thus, we have created a rigorous proof that the transitivity property  can be violated for statements about measurements. 
Note that this violation occurs for a microscopic system 
since spins are associated with microscopic objects. 
No measurements on the agents themselves are involved. 

In logic, one can consider the situation in which two premises P and Q must be both satisfied.
This is logical conjunction and is denoted by (P AND Q). 
For example, in Eq.(\ref{InitialStateLogic}), 
if P = (barred spin is up) and Q = (unbarred spin is down), 
then (P AND Q) means that both conditions hold and one is restricting the situation to the first term in Eq.(\ref{InitialStateLogic}).
This is a valid use of conjuction in quantum mechanics. 

Now consider, P = (unbarred spin is $\uspin_z$) and Q = (unbarred spin is ${(\uspin_z - \dspin_z)/\sqrt{2}}$).\footnote{We 
choose a minus sign in Q because it corresponds closer to the situation in reference \refcite{FRGE}.
}
These premises are incompatible and using them with conjuction is an invalid operation. 
One might think that (P AND Q) means (unbarred spin is  $\uspin_z$ since $\uspin_z$ is common to both P and Q. 
However, if one uses the $x$ axis of quantization, 
then P = (unbarred spin is ${(\uspin_x + \dspin_x)/\sqrt{2}}$) and Q = (unbarred spin is $\dspin_x$), 
and using the same faulty reasoning one might conclude that \mbox{(P AND Q)} means  (unbarred spin is  $\dspin_x$).
In short, there is no way to define (P AND Q) for this situation. 

A relevant generalization, which involves measurements, is the following: 
Suppose the wavefunction at time $t_0$ is
\begin{equation}
     \Psi_0 =   
   \frac{\Psi^+_0}{\sqrt{2}} (\Psi^F_{\uparrow} + \Psi^F_{\downarrow}) + \frac{\Psi^-_0}{\sqrt{2}} (\Psi^F_{\uparrow} - \Psi^F_{\downarrow})
\ ,
\label{ConjunctionIssue0} 
\end{equation}
where $\Psi^+_0$ and $\Psi^-_0$ involve the barred spin but not the unbarred spin, and $|\Psi^+_0|^2 + |\Psi^-_0|^2 = 1$.
Suppose that the barred and unbarred degrees of freedom evolve independently. 
Then, the evolution of $\Psi_0$ maintains its factorized form, 
that is, at time $t$, 
\begin{equation}
     \Psi_t =  \Psi^+_t \Psi^W_{+} + \Psi^-_t \Psi^W_{-}
\ ,
\label{ConjunctionIssuet} 
\end{equation}
where $\Psi^+_0$ evolves to $\Psi^+_t$,  ${{(\uspin_z + \dspin_z)}/{\sqrt{2}}}$ evolves to $ \Psi^W_{+}$, 
$\Psi^-_0$ evolves to $ \Psi^-_t$, and ${{(\uspin_z - \dspin_z)}/{\sqrt{2}}}$ evolves to $ \Psi^W_{-}$.
Suppose P = (Agent F measures unbarred spin to be up at time $t_0$) and Q = (the measurement of Agent W is `$-$' at time $t$).  
Because of factorized evolution, 
premise Q implies that the relevant part of the wavefunction is the second term in Eq.(\ref{ConjunctionIssue0}); 
it is proportional to $\Psi^F_{\uparrow}-\Psi^F_{\downarrow}$ 
and incompatible with premise P, which says that Agent F measured the spin to be up at time $t_0$.
Hence, (P AND Q) has no logical sense. 
In the last part of Section \ref{IssuesWithLogic}, 
we show that certain pairs of premises in the Frauchiger-Renner argument 
are incompatible with logical conjunction.

At the beginning of this section, 
we promised to show that Mathematical Statements 1-4 can individually be verified experimentally.
To perform this task, we introduce two ``Referee Agents'' R and  $\bar{\mathrm{R}}$.
The initial wavefunction is similar to Eqs.(\ref{InitialState}) and (\ref{InitialStateLogic}): 
\begin{equation}
     \Psi =    \frac{\Psi^R \psibar{R} }{\sqrt{3}} 
        (\uspinbar_z \dspin_z + \dspinbar_z \uspin_z + \dspinbar_z \dspin_z)
\ .
\label{InitialStateVerificationzz} 
\end{equation}
Mathematical Statement 1 in Eq.(\ref{LogicStatement1}) is automatically true. 
To verify Mathematical Statement 3 (Eq.(\ref{LogicStatement3})), 
Referee Agents R and  $\bar{\mathrm{R}}$ make spin measurements using the $z$-axis for spin quantization. 
The wavefunction becomes
\begin{equation}
     \Psi' =    \frac{1}{\sqrt{3}} 
        (\psibar{R}_{\uparrow_z} \Psi^{R}_{\downarrow_z} + \psibar{R}_{\downarrow_z} \Psi^{R}_{\uparrow_z} + \psibar{R}_{\downarrow_z} \Psi^{R}_{\downarrow_z})
\ .
\label{FinalStateVerificationzz} 
\end{equation}
Then, Agents R and $\bar{\mathrm{R}}$ get together to compare results. 
Whenever Agent R measures ${\uparrow_z}$, he finds that Agent $\bar{\mathrm{R}}$ measures $\bar{\downarrow}_z$ 
due to the middle term in Eq.(\ref{FinalStateVerificationzz}) and the lack of a $\psibar{R}_{\uparrow_z} \Psi^{R}_{\uparrow_z}$ term. 

To verify Mathematical Statement 2 (Eq.(\ref{LogicStatement2})), 
Referee Agent R uses the $z$-axis to make the spin measurement while Agent $\bar{\mathrm{R}}$ uses the $x$-axis. 
Using these bases, the initial state in Eq.(\ref{InitialStateVerificationzz}) becomes
\begin{equation}
\begin{aligned}
     \Psi =    \frac{\Psi^R \psibar{R} }{\sqrt{6}} &
        ((\uspinbar_x + \dspinbar_x) \dspin_z + (\uspinbar_x - \dspinbar_x) \uspin_z + (\uspinbar_x - \dspinbar_x) \dspin_z) \\
           =  &  \frac{\Psi^R \psibar{R} }{\sqrt{6}}  (2\uspinbar_x  \dspin_z + \uspinbar_x \uspin_z - \dspinbar_x \uspin_z)
\ .
\label{InitialStateVerificationzx} 
\end{aligned}
\end{equation}
After the measurements are made, the wavefunction becomes
\begin{equation}
     \Psi' =  \frac{1}{\sqrt{6}}  (2 \psibar{R}_{\uparrow_x}  \Psi^{R}_{\downarrow_z} +  \psibar{R}_{\uparrow_x} \Psi^{R}_{\uparrow_z} - 
            \psibar{R}_{\downarrow_x} \Psi^{R}_{\uparrow_z})
\ .
\label{FinalStateVerificationzx} 
\end{equation}
When  Agents R and $\bar{\mathrm{R}}$ get together, 
they find that whenever Agent $\bar{\mathrm{R}}$ measures $\bar{\downarrow}_x$,  Agent R measures ${\uparrow_z}$.
This is because of the last term in Eq.(\ref{FinalStateVerificationzx}) and the lack of a $\psibar{R}_{\downarrow_x} \Psi^{R}_{\downarrow_z}$ term.

To verify Mathematical Statement 4 (Eq.(\ref{LogicStatement4})), 
Agent R uses the $x$-axis to make the spin measurement while Agent $\bar{\mathrm{R}}$ uses the $z$-axis. 
Using these bases, the initial state in Eq.(\ref{InitialStateVerificationzz}) becomes
\begin{equation}
\begin{aligned}
     \Psi =    \frac{\Psi^R \psibar{R} }{\sqrt{6}} &
        (\uspinbar_z (\uspin_x - \dspin_x) + \dspinbar_z (\uspin_x + \dspin_x) + \dspinbar_z (\uspin_x - \dspin_x)) \\
           =  &  \frac{\Psi^R \psibar{R} }{\sqrt{6}}  (\uspinbar_z \uspin_x - \uspinbar_z  \dspin_x  + 2  \dspinbar_z  \uspin_x)
\ .
\label{InitialStateVerificationxz} 
\end{aligned}
\end{equation}
After the measurements are made, the wavefunction becomes
\begin{equation}
     \Psi' =  \frac{1}{\sqrt{6}}  (\psibar{R}_{\uparrow_z} \Psi^{R}_{\uparrow_x} - \psibar{R}_{\uparrow_z}  \Psi^{R}_{\downarrow_x} 
            + 2\psibar{R}_{\downarrow_z} \Psi^{R}_{\uparrow_x})
\ .
\label{FinalStateVerificationxz} 
\end{equation}
When  Agents R and $\bar{\mathrm{R}}$ get together, 
they find that whenever Agent $\bar{\mathrm{R}}$ measures $\bar{\downarrow}_z$,  Agent R measures ${\uparrow_x}$.
This is because of the last term in Eq.(\ref{FinalStateVerificationxz}).

Physicists might argue that mathematical statements about a wavefunction of the type discussed in this section 
are not physical because they are not observed. 
However, we believe it is more reasonable to say that 
mathematical statements about a wavefunction are physcially valid 
if there {\it exists a measurement} that can show that they are true. 
With this viewpoint, Mathematical Statements 1 - 4 are individually physically true statements.  

Suppose that one attempts to use in {\it single} measurement procedure -- similar to the above technique of using referee agents -- to show 
the validity of any two of the four Mathematical Statements. 
Then, one finds that this cannot be done. 
A measurement to establish the validity of one of the statements interferes with the measurement to establish the validity of the other statement, or vice versa. 
In the case of Mathematical Statements 1 - 4, 
one runs into the issue of compatibility of premises 
as explained above (Discussion starts at two paragraphs above Eq.(\ref{ConjunctionIssue0})). 
One would have to measure either the barred spin or the unbarred spin in both the $z$-direction and the $x$-direction.
Also problematic is that after a measurement, 
the quantum degrees of an agent become entangled with the spin that he measures. 
Therefore, the second of the two measurements would have to be performed on the degrees of freedom of the agent  
and this can be problematic; see Section \ref{WignerMeasurement}.
The discussion here can be considered as providing an understanding 
of why classical logic cannot always be applied when combining certain ``If ..., then ...'' statements in quantum mechanics. 

It one were to attempt to carry out the Frauchiger-Renner experiment in the real world, 
then one would have to make sure that the correct initial spin state of Eq.(\ref{InitialState}) 
is being generated. 
The way to do this is to perform measurements on the spins, 
and this turns out to be the same procedure used to experimentally verify Mathematical Statements 2-4;
see Eqs.(\ref{InitialStateVerificationzz}) - (\ref{FinalStateVerificationxz}) above and the corresponding discussion.
For example, the verification of Statement 3 above shows 
that the spin state must contain the term $\dspinbar_z \uspin_z$ but cannot contain $\uspinbar_z \uspin_z$. 
It can be checked that the verification of Statements 2 - 4 using measurements is sufficient to uniquely determine 
the spin wavefunction to be that in Eq.(\ref{FinalStateVerificationzz}). 
In a real experiment, one would repeatedly check the intial spin-state generation until one gained confidence 
that one is producing it correctly each time. 

One can also use referee agents to experimentally verify the statements in the Frauchiger-Renner gedanken experiments.
As above, we need two of them: Agents R and $\bar{\mathrm{R}}$.
The initial wavefunction in Eq.(\ref{InitialState}) is replace by
 \begin{equation}
     \Psi_0 =    \frac{\Psi^R \psibar{R} \Psi^W \psibar{W} \Psi^F \psibar{F} }{\sqrt{3}} 
        (\uspinbar \dspin + \dspinbar \uspin + \dspinbar \dspin)
\ .
\label{InitialStateRefereeing} 
\end{equation}
To experimentally verify Statement 2, namely, if Agent $\bar{\mathrm{W}}$ measures `$\overline{-}$' at time $t_3$, then Agent F measured 
the unbarred spin to be up at time $t_2$,
one has Agent R {\it observe} Agent F just after time $t_2$.
Refer to footnote c above.
An observation of Agent R on Agent F's wavefunction takes the form: 
\begin{equation}
\begin{aligned}
\Psi^R \uspin_M  & \to \Psi^R_\uparrow \uspin_M \\
\Psi^R \dspin_M  & \to \Psi^R_\downarrow \dspin_M  
\ ,
\label{RobservingF} 
\end{aligned}
\end{equation}
so that Agent R does not disturb the wavefunction of Agent F. 
The wavefunction just after $t_2$ takes the form
\begin{equation}
     \Psi'_2 =   \frac{\psibar{R} \Psi^W  \psibar{W} }{\sqrt{3}} 
         (\Psi^R_\downarrow \uspinbar_M  \dspin_M + \Psi^R_\uparrow \dspinbar_M \uspin_M +  \Psi^R_\downarrow \dspinbar_M \dspin_M)
\ .
\label{StateTime2o} 
\end{equation}
Then, at a time just after $t_3$, 
Agent $\bar{\mathrm{R}}$ observes what Agent  $\bar{\mathrm{W}}$ has measured. 
The analog of Eq.(\ref{RobservingF}) is 
\begin{equation}
\begin{aligned}
\psibar{R} \psibar{W}_+  & \to \psibar{R}_+ \psibar{W}_+ \\
\psibar{R} \psibar{W}_-  & \to \psibar{R}_- \psibar{W}_-  
\ ,
\label{RBARobservingWBAR} 
\end{aligned}
\end{equation}
and the wavefunction just after $t_3$ becomes
\begin{equation}
\begin{aligned}
     \Psi'_3 = & \frac{\Psi^W}{\sqrt{6}} 
          {\big( \Psi^R_\downarrow (  \psibar{R}_+ \psibar{W}_+ + \psibar{R}_- \psibar{W}_-) \dspin_M }\\
    + \  & { \Psi^R_\uparrow ( \psibar{R}_+ \psibar{W}_+ - \psibar{R}_- \psibar{W}_-) \uspin_M + \Psi^R_\downarrow ( \psibar{R}_+ \psibar{W}_+ - \psibar{R}_-  \psibar{W}_-)   \dspin_M \big)} = \\ 
                 & \frac{\Psi^W}{\sqrt{6}}  
            {\big( 2 \psibar{R}_+ \Psi^R_\downarrow \psibar{W}_+  \dspin_M  +  \psibar{R}_+  \Psi^R_\uparrow \psibar{W}_+  \uspin_M - \psibar{R}_-  \Psi^R_\uparrow \psibar{W}_- \uspin_M  \big)}
\ .
\label{StateTime4o} 
\end{aligned}
\end{equation}
Next, Agents R and $\bar{\mathrm{R}}$ meet. 
They find that whenever Agent $\bar{\mathrm{R}}$ observed Agent $\bar{\mathrm{W}}$ to measure `$\overline{-}$', 
then Agent R had to observe Agent F measuring $\uspin$. 
This result comes from the last term in Eq.(\ref{StateTime4o}).

In a similar manner, Statement 3 is experimentally verified by having Agent $\bar{\mathrm{R}}$ observe 
Agent $\bar{\mathrm{F}}$'s measurement just after time $t_1$ and then 
having Agent R observe Agent F's measurement just after time $t_2$.
Statement 4 is verified by having Agent $\bar{\mathrm{R}}$ observe Agent $\bar{\mathrm{F}}$'s measurement  just after time $t_1$ and then 
having Agent R observe Agent W's measurement just after time $t_4$. 
Although Statement 1$'$ is obviously true, 
one could have agents $\bar{\mathrm{R}}$ and R respectively observe the measurements of $\bar{\mathrm{W}}$ and agents W 
just after times $t_3$ and $t_4$ respectively.

\section{Quantum Logic and Measurement Statements}
\label{MeasurementLogic}

The example associated with Eqs.(\ref{InitialState23}) - (\ref{State23Time2}) above involves going ``backward and forward'' in time. 
Is there an example in which one avoids this? 
Consider a system involving three spin-$\oneHalf$ objects and three agents A, B and C who perform measurements on them. 
Use the following as the initial wavefunction: 
\begin{equation}
     \Psi_0 =     \Psi^A \Psi^B \Psi^C 
        (\frac{1}{\sqrt{2}}\uspin_A \uspin_B \uspin_C + \frac{1}{\sqrt{2}}\dspin_A \uspin_B \dspin_C )
\ . 
\label{SimplerExample}
\end{equation}
Agent A first measures A-spin at time $t_A$, then Agent B measures B-spin at time $t_B$ and 
finally Agent C measures C-spin at $t_C$. Here, $t_A < t_B < t_C$.
After these measurements are made the structure of the wavefunction is 
\begin{equation}
     \Psi_f =  
        \frac{1}{\sqrt{2}} \Psi^A_\uparrow \Psi^B_\uparrow \Psi^C_\uparrow + \frac{1}{\sqrt{2}} \Psi^A_\downarrow \Psi^B_\uparrow \Psi^C_\downarrow 
\ . 
\label{SimplerExampleFinal}
\end{equation}
The agents A, B, and C can then get together to discuss their results. 
The following statements are verified to be true from Eq.(\ref{SimplerExampleFinal}) 
using the quantum logic rules (i)-(iii): 
$$
\begin{aligned}
   \textbf{Statement A:}&\textrm{ If Agent A measures A-spin to be up at time }t_A , \\
                                  &\textrm{\ \ \ \ \ \  then Agent B will measure B-spin to be up at time }t_B. \\
   \textbf{Statement T:}&\textrm{ If Agent A measures A-spin to be up at time }t_A , \\
                                  &\textrm{\ \ \ \ \ \  then Agent C will measure C-spin to be up at time }t_C. 
\end{aligned}
$$
One also can deduce: 
$$
\begin{aligned}
  &\textbf{Statement B:}\textrm{ If Agent B measures B-spin to be up at time }t_B , \\
  &\textrm{\ \ \ \ then Agent C at time }t_C\textrm{ will not necessarily measure C-spin to be up.} 
\end{aligned}
$$
Indeed, if Agent B measures B-spin to be up at time $t_B$, 
then Agent C will measure C-spin to be up 50\% of the time and down 50\% of the time.
If Statements A and B could be combined using the transitive property of logic, then one would obtain 
$$
\begin{aligned}
  &\textbf{Statement F:}\textrm{ If Agent A measures A-spin to be up at time }t_A , \textrm{ then}\\
  &\textrm{\ \ it is not guaranteed that Agent C at time }t_C\textrm{ will measure C-spin to be up.}
\label{SimplerExampleFalseStatement}
\end{aligned}
$$
Statement F is false since it violates Statement T, the latter always being true. 
Therefore, the transitive property of logic can be violated in quantum mechanics 
concerning statements about measurements. 
Again, this is for a microscopic system since spins are involved.
If there is any doubt about the above, Statements A, B and T can be verified in a real experiment; 
the difficult part is in generating the initial entangled spin state, but nowadays there are methods to handle this. 

In standard logic, an ``If ... then ...'' statement cannot be ``50\% true''; 
if it is not always true, then it is considered false. 
However, given that the conclusion of the ``If ... then ...'' Statement F (obtained using transitivity) and that Statement B 
can be replaced by ``Agent C will measure C-spin to be up 50\% of the time and down 50\% of the time'',
one might characterize Statement F as being ``50\% true'' and ``50\%' false''.\footnote{By performing a series of runs 
and collecting statistics, it can be verified that Statement F produces a false result among ``50\%'' of the runs. 
It should be clear that standard classical logic is not the correct framework 
for dealing with statements about wavefunctions and measurements. 
In fuzzy logic,\cite{FuzzyLogin} it is permissible to have statements that are ``fractionally'' true.}
If in the above example, 
the coefficient of $\uspin_A \uspin_B \uspin_C$ in $\Psi_0$ is selected to be $\sqrt{0.1}$ 
while that of $\dspin_A \uspin_B \dspin_C $ is $\sqrt{0.9}$, 
then Statement F is ``90\% false'' and ``10\% true''. 
One can adjust the component coefficients to ``increase'' the ``falsehood'' of F, 
but one cannot arrive at ``100\%'' in this simple example. 
In addition, up to this point,
all of the examples of violations of transitivity for statements about measurements 
involve using an ``If ... then ..." statement in which one of the conclusions of a premise involves probabilities; 
it is quite natural and acceptable to have such statements since quantum mechanics involves uncertainty and probabilities. 
In the Frauchiger-Renner gedanken experiment, 
none of conlcusion of Statements 1 - 4 involve probabilities; 
however, in the next section, we show that violations of transitivity and other rules of logic still arise. 

Suppose that we have a wavefunction at time $t=0$ of the form  
\begin{equation}
\Psi_0 = c_a \Psi^a_0 +  c_b \Psi^b_0 
\ ,
\label{HypotheticalWavefunction}
\end{equation}
where $\Psi^a_0$ and $\Psi^b_0$ are orthogonal normal and $c_a$ and $c_b$ are constants.
Suppose that one is trying to combine two ``If ... then ...'' statements using transitivity 
but the premise of the first ``If ... then ...'' statement involves a premise restricting the wavefunction to $  c_a \Psi^a_0$ but a conclusion that involves both terms, 
while the premise of the second ``If ... then ...'' statement is this conclusion (and hence involves both terms). 
This is the case in the above example. 
It is accomplished by a ``shift effect'': 
$\Psi_0 = c_a \Psi^a_0 +  c_b \Psi^b_0 = c_a \uspin_A \uspin_B  \uspin_C + c_b \dspin_A \uspin_B \dspin_C$.
The premise of ``If $\uspin_A$ at time $t_A$, then $\uspin_B$ at time $t_B$'' 
involves only the first term of $\Psi_0$ (that is, the $c_a \Psi^a_0 = c_a \uspin_A \uspin_B \uspin_C $ term), 
but the premise of ``If $\uspin_B$ at time $t_B$, then ...'' involves both terms. 
Roughly speaking, the premise has ``shifted'' from the first term to both terms. 
The statement ``If $\uspin_B$ at time $t_B$, then ... '' 
can be true because of the second term ($c_b\dspin_A \uspin_B \dspin_C$)
and this can mean that the premise ``If $\uspin_A$ at time $t_A$''  
of ``If $\uspin_A$ at time $t_A$, then $\uspin_B$'' can be false. 
Indeed, this is the origin of the violations of transitivity in the examples presented so far. 
Consider Statements A and B above. 
When the premise of B is true but the premise of A is false, 
A-spin is down and this corresponds exactly to the cases in which transitivity 
leads to a false result (that is, C-spin is down) in Statement F.\footnote{This particular understanding of 
the violation of transitivity grew out of email exchanges that I had with Renato Renner 
from May 1, 2022 to May 13, 2022. 
}

Unitarity guarantees that the evolutions of $\Psi^a_0$ and $\Psi^b_0$ 
to a later time $t$ (which we denote by $\Psi^a_t$ and $\Psi^b_t$) proceeds independently 
and that the two components remain orthogonal normal. 
This means that the future evolution of  $\Psi^a_t$ cannot depend on $\Psi^b_t$ and vice versa. 
If the conclusions in the two ``If ... then ...'' statements are only ``compatible'' due to the first term in the wavefunction,  
then the use of transitivity 
will be violated in a fraction of the cases given by: 
\begin{equation}
 \textrm{violation fraction} = {{|c_a|^2} \over {|c_a|^2+|c_b|^2}}
\ .
\label{ViolationFraction}
\end{equation}
Since in the above examples $c_a = c_b$, the ``violation of transitivity is 50\%''. 
It is possble for the conclusions for the two ``If ... then ...'' statements to be ``compatible'' for both terms in the wavefunction, 
in which case  transitivity produces a valid statement, 
even when the ``shift effect'' is present. 
This can be considered a coincidence. 
For example, 
if the C-spin of the second term in Eq.(\ref{SimplerExample}) is changed to be up, 
then the conclusions of statement B and F above are both changed to ``Agent C at time $t_C$ will measure C-spin to be up''. 
However, the reason that Statement F is now true is because Agent C measures C-spin to be up for both terms. 
In mathematical statements about wavefunctions, the ``shift effect'' can also be traced to the reason why 
the incorrect result in Eq.(\ref{LogicStatement4m}) arises when one makes use of logical transitivity. 

In the next section, we show that the combining of Statements 1$'$ and 2, of Statements 2 and 3 and of Statements 3 and 4 
using transitivity 
in the Frauchiger-Renner gedanken experiment uses a ``shift effect'' 
and involves exactly the same structure discussed here with $c_a = c_b$. 
Hence, the logical statements obtained from them using transitivity are not true.

In addition to transitivity, there are other rules of logic that are violated in quantum mechanics. 
For example, in standard classical logic, 
If $P$ implies $R$, and $Q$ is any other condition, then ($P$ and $Q$) also implies $R$:
$(P \Rightarrow R) \Rightarrow ((P \ \textrm{AND} \ Q) \Rightarrow R)$.
Return to the experiment associated with Eqs.(\ref{SimplerExample}) and (\ref{SimplerExampleFinal}), 
and let $P$ to be the premise of Statement B ($P =$ ``Agent B measures B-spin to be up at time $t_B$''), 
let $R$ be the conclusion of Statement B ($R =$ ``Agent C at time $t_C$ will not necessarily measure C-spin to be up''), 
and let $Q$ be the premise of Statement A ($Q =$ ``Agent~A measures A-spin to be up at time $t_A$'').
Then P AND Q actually implies S instead of $R$;  
the conclusion $S$ is ``Agent C at time $t_C$ will measure C-spin to be up''. 

We now illustrate the power of unitary by showing that 
the four measurement statements of the Frauchiger-Renner gedanken experiment 
can all be derived from the final wavefunction in Eq.(\ref{StateTimeFinal}) 
and knowledge of how the experiment was conducted, that is, 
Agent $\bar{\mathrm{F}}$ first measured barred spin at time $t_1$, then Agent F  measured unbarred spin at time $t_2$, etc.
Note that $\uspin_M$ (respectively, $\dspin_M$) at time $t_2$ always evolves to 
$(\Psi^W_+ + \Psi^W_-)/\sqrt{2}$ (respectively, $(\Psi^W_+ - \Psi^W_-)/\sqrt{2}$) at time $t_4$. 
The analogous statement is true for the barred states. 

The first step is to rewrite Eq.(\ref{StateTimeFinal}) so that the $(\Psi^W_+ +  \Psi^W_-)$ 
and $(\Psi^W_+ -  \Psi^W_-)$ dependence is evident: 
\begin{equation}
     \Psi_4 = \frac{1}{\sqrt{12}}  \big(  (2\psibar{W}_+ ( \Psi^W_+ -  \Psi^W_-) 
          +   (\psibar{W}_+ - \psibar{W}_-)  ( \Psi^W_+ +  \Psi^W_-)   
   \big)
\ . 
\label{StateTime4A} 
\end{equation}
Statement 1 is a tautology: ``If agents W and $\bar{\mathrm{W}}$ respectively measured 
`$-$' and `$\overline{-}$', then $\bar{\mathrm{W}}$ measured `$\overline{-}$'.'' 
Now look at what multiplies $\psibar{W}_-$ in Eq.(\ref{StateTime4A}).
It is ${(\Psi^W_+ + \Psi^W_-)}$. 
Since it had to have evolved from $\uspin_M$, one derives  
Statement 2: ``If Agent $\bar{\mathrm{W}}$ obtained a measurement of `$\overline{-}$' at time $t_3$, 
then Agent F previously measured unbarred spin to be up.''
Next look at the factor that multiplies what evolves from $\uspin_M$, namely ${(\Psi^W_+ + \Psi^W_-)}$.
This factor is $ (\psibar{W}_+ - \psibar{W}_-)$ and evolved from $\dspinbar_M$. 
Hence, one obtains Statement 3: ``If Agent F measured unbarred spin up, 
then Agent $\bar{\mathrm{F}}$ measured barred spin to be down.''

To obtain Statement 4, one needs to rewrite Eq.(\ref{StateTimeFinal}) so that the $(\psibar{W}_+ + \psibar{W}_-)$ 
and $(\psibar{W}_+ - \psibar{W}_-)$ dependence is evident: 
\begin{equation}
     \Psi_4 = \frac{1}{\sqrt{12}}  \big(  (2  (\psibar{W}_+ - \psibar{W}_-)  \Psi^W_+ 
          +   (\psibar{W}_+ + \psibar{W}_-)  ( \Psi^W_+ -  \Psi^W_-)   
   \big)
\ . 
\label{StateTime4B} 
\end{equation}
Barred down spin evolves to $ (\psibar{W}_+ - \psibar{W}_-)$ and 
it multiplies $ \Psi^W_+$ (the first term in Eq.(\ref{StateTime4B})), 
and so one obtains 
``If Agent $\bar{\mathrm{F}}$ previously measured barred spin down, then 
Agent W will obtain `$+$' for his measurement,'' 
which is Statement 4. 
The above shows that there is a close relation with the measurement statements in the Frauchiger-Renner gedanken experiment 
and the mathematical statements of Section \ref{WavefunctionLogic}, 
particularly those in the paragraphs above and below Eq.(\ref{RotationBy90p}), 
and that the reasons for the violation of transitivity are similar. 

\section{Issues with Logic in the Frauchiger-Renner Argument }
\label{IssuesWithLogic}

Agent F can perform her measurement before Agent $\bar{\mathrm{F}}$ 
and the resulting wavefunction after both measurements remains the same. 
One can have $t_1$ approach $t_2$ or even have the two agents perform their measurements at the same time.\footnote{As an aside, 
it is also true that the temporal order does not matter for the two Wigner measurements at times  $t_3$ and $t_4$;  
one can have $t_3 < t_4$, $t_4 < t_3$ or $t_3 = t_4$, 
and Statements 1-4 of the Frauchiger-Renner gedanken experiment are all still valid.}
Then Statements 3 and 4 are valid at the same time. 
So, one can take $t_1 = t_2$ and use Eq.(\ref{StateTime2}) as the starting point for the Frauchiger-Renner argument.
Below, we often make this simplification. 

Consider the logic involved in combining Statement 3 and Statement 4 
in the logical chain that leads to the Contradictory Statement. 
Let $P$ be the premise of Statement 3, that is, 
$P =$ (Agent F measured the unbarred spin to be up at time $t_2$).
Let $Q$ be the conclusion of Statement 3, which is also the premise of Statement 4. 
Here, $Q =$ (Agent $\bar{\mathrm{F}}$ measured the barred spin to be down at time $t_1 = t_2$). 
Finally, let $R$ be the conclusion of Statement 4:  
$R =$ (Agent W will measure `$+$' at time $t_4$).
Statement 3 is $P \Rightarrow Q$ and Statement 4 is $Q \Rightarrow R$. 
Now, the premise of Statement 4 involves the second and third terms in the wavefunction at time $t_2$ 
(those involving $\dspinbar_M \uspin_M + \dspinbar_M \dspin_M$ in Eq.(\ref{StateTime2})) 
while Statement 3 involves the middle or second term (the one involving $\dspinbar_M \uspin_M$). 
Let us just focus of this part of the wavefunction and its evolution to time $t_4$:
\begin{equation}
\begin{aligned}
     \Psi_2 =   \frac{\Psi^W  \psibar{W} \sqrt{2} }{\sqrt{3}} 
         &\big( ... + \sqrt{1 \over 2} \dspinbar_M \uspin_M + \sqrt{1 \over 2} \dspinbar_M \dspin_M \big) \\ 
         &\ \ \ \ \ \ \ \ \ \ \ \ \ \ \ \ \ \ \ \ \ \ \ \ \ \ \ . \\
         &\ \ \ \ \ \ \ \ \ \ \ \ \ \ \ \ \ \ \ \ \ \ \ \ \ \ \ . \\
     \Psi_4 =   \sqrt{\frac{2}{3}} 
        \big(  ...&  + \frac{1}{2}\psibar{W}_{\downarrow} ( \Psi^W_+ +  \Psi^W_-)  + \frac{1}{2}\psibar{W}_{\downarrow} ( \Psi^W_+ -  \Psi^W_-)   \big)
\ ,
\label{States23Time2} 
\end{aligned}
\end{equation}
where $\psibar{W}_{\downarrow}$ is an abbreviation for $(\psibar{W}_+ - \psibar{W}_-)/\sqrt{2}$.
The first thing to note is that the ``shift effect'' is occurring, 
and so, given the results in Section \ref{MeasurementLogic}, 
combining Statements 3 and 4 using transitivity is an invalid procedure. 
One can also ``quantify'' the violation of assuming 
${(P \Rightarrow Q) \ \textrm{AND} \ (Q \Rightarrow R))\Rightarrow (P \Rightarrow R)}$
using Eq.{\ref{ViolationFraction}): $P \Rightarrow R$ should be ``50\% false'', 
which is easy to show. 

The $\dspinbar_M \uspin_M$ term in Eq.(\ref{States23Time2}) at time $t_2$ evolves 
to the $\psibar{W}_{\downarrow} ( \Psi^W_+ +  \Psi^W_-)$ term at time $t_4$.
Likewise, the last term in Eq.(\ref{States23Time2}) at $t_2$ evolves to the last term at $t_4$ in $\Psi_4$.
When the premise $Q$ (i.e., barred spin is measured to be down) of Statement 4 holds, 
both terms are relevant and a cancellation of $\Psi^W_-$ occurs in $\Psi_4$ thereby yielding the conclusion $R$ of Statement 4, 
that the probability of Agent W obtaining  `$+$' is 100\%.
However, if an ``If ... then ...'' statement involves premise $P$ (i.e., unbarred spin is measured to be up), 
then the relevant term is the $\dspinbar_M \uspin_M$ one. 
It evolves to something proportional to $ ( \Psi^W_+ +  \Psi^W_-)$.
So, if $P$ holds, that is, Agent F measured unbarred spin up at time $t_2$, 
then there is a 50\% chance Agent W will obtain `$+$' for his measurement and not 100\%. 
Hence, this`` direct'' calculation demonstrates that the transitivity property of logic 
cannot be used to combine Statements 3 and 4; 
its use produces an incorrect result. 
The different understandings of why transitivity is violated in the Frauchiger-Renner case 
are identical to those of why transitivity is violated for the ``A-B-C'' model of the previous section.
However, for  the Frauchiger-Renner case, there is another explanation.

\begin{figure}[h]
\centerline{\includegraphics[width=11.9cm, height=8.275cm]{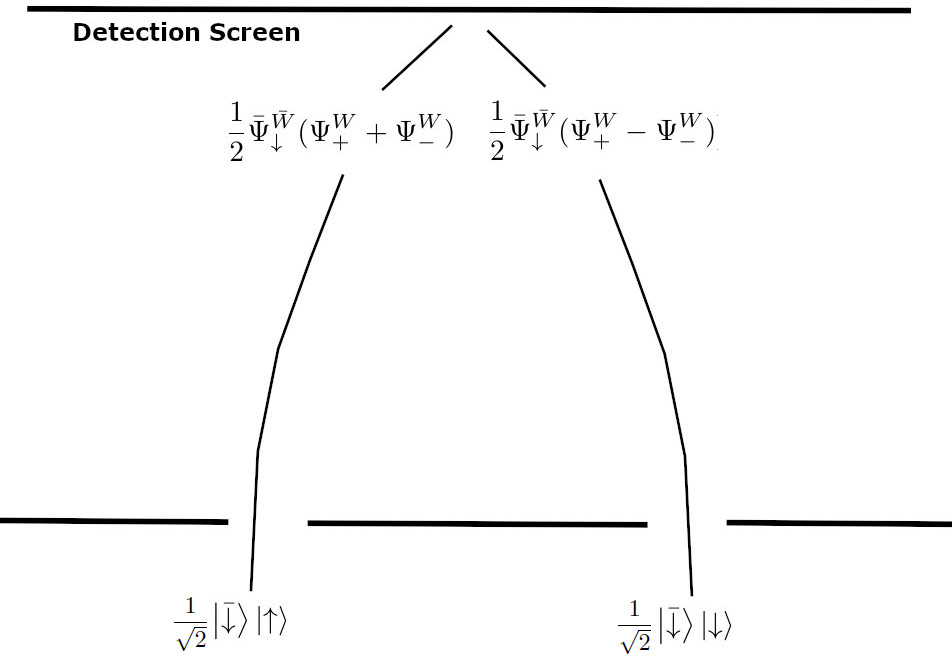}}
\vspace*{1pt}
\caption{Interference Effect Involved in Statement 4.}
\label{fig:Inteference}
\end{figure}

The conclusion of Statement 4 involves a delicate cancellation between two terms in Eq.(\ref{States23Time2}) 
to eliminate the $\Psi^W_-$ dependence in $\Psi_4$.
This can be considered a quantum interference effect. 
The best example of quantum interference is the double slit experiment. 
Imagine that $\dspinbar \uspin/\sqrt{2}$ is associated with passing through the left slit, 
while $\dspinbar \dspin/\sqrt{2}$ is associated with passing through the right slit. 
Each evolves respectively to ${\psibar{W}_{\downarrow} ( \Psi^W_+ +  \Psi^W_-)/2}$ 
and ${\psibar{W}_{\downarrow} ( \Psi^W_+ -  \Psi^W_-)/2}$ 
when they reach a certain point on the detection screen. 
See  Figure \ref{fig:Inteference}. 
If no attempt is made to detect whether the wavefunction goes through the left or right slit, 
then the quantum interference effect occurs, there is a ``cancellation of the $ \Psi^W_-$ amplitude'', 
and the screen will signal to Agent W a `$+$' outcome. 
Now when premise $P$ is operative, it means that Agent F has effectively ``done something'' 
to determine which slit the object went through. 
Indeed, she has determined that it went through the left slit 
because she has measured the unbarred spin to be up, 
which is associated with $\dspinbar \uspin/\sqrt{2}$. 
This disturbs the quantum interference effect, 
the wavefunction will evolve to $\psibar{W}_{\downarrow} ( \Psi^W_+ +  \Psi^W_-)/2$  at the screen,  
and the signal can no longer be guaranteed to be `$+$'. 
Half the time it will be `$+$' and half the time it will be `$-$'.
Thus, using the analogy with the two-slit experiment, 
one understands physically why the transitive rule is violated in this case: 
When Statement 3 is combined with Statement 4, 
the resulting logical statement does not properly take into account the effect 
of the measurement performed by Agent F on the measurement by Agent W. 

If the gedanken experiment involved only the terms displayed in Eq.(\ref{States23Time2}), 
then one can derive the following two statements: 

{\bf Statement 3L}: If Agent F measures the unbarred spin to be up at time $t_2$, 
then Agent W at time $t_4$ will measure `$+$' 50\% of the time. 

{\bf Statement 3R}:  If Agent F measures the unbarred spin to be down at time $t_2$, 
then Agent W at time $t_4$ will measure `$+$' 50\% of the time. \\
Now, in classical logic, if ${(L \Rightarrow Q)}$ and ${(R \Rightarrow Q)}$ 
are two valid ``If ... then ...'' statements, 
then one can conclude that ${(L \ \textrm{OR} \ R) \Rightarrow Q}$. 
Here, $L$ and $R$ are respectively the premises of Statements 3L and 3R, 
and $Q =$ (Agent W at time $t_4$ will measure `$+$' 50\% of the time), 
or one can use $Q =$ (Agent W at time $t_4$ will not necessarily obtain a measurement of `$+$').
However, the correct statement involving the premise $(L \ \textrm{OR} \ R )$ 
is  $(L \ \textrm{OR} \ R ) \Rightarrow Q'$, 
where $Q' =$ (Agent W at time $t_4$ will measure `$+$' with certainty).
This is just another example of how classical logic cannot be applied to statements about measurements, 
especially when quantum interference effects are involved. 

When Statement 2 is combined with Statement 3 using transitivity, 
the result is `` If Agent $\bar{\mathrm{W}}$ measured `$\overline{-}$' at time $t_3$, 
then Agent $\bar{\mathrm{F}}$ measured the barred spin to be down ($\dspinbar$) at time $t_1$.
However, we know from the methods used in the last two paragraphs of Section \ref{MeasurementLogic} that 
that if Agent $\bar{\mathrm{W}}$ measured `$\overline{-}$' at time $t_3$, 
then it had to have evolved from a term proportional to $\uspinbar_M - \dspinbar_M$ in $\Psi_1$ at time $t_1$ 
and not something proportion to $\dspinbar_M$.
In fact, one can show that it originated from $(\uspinbar_M - \dspinbar_M) \uspin$ (up to a factor).
Hence, combining Statement 2 with Statement 3 using transitivity generates an invalid logic statement. 
It is violated 50\% of the time. 

To analyze whether Statement 1$'$ can be combined with Statement 2 using transitivity, 
one needs to consider the last two terms in $\Psi_4$ of Eq.(\ref{StateTimeFinal}).
Recall that Statement 1$'$ is 
``If, at time $t_4$, agents W and $\bar{\mathrm{W}}$ respectively measure 
`$-$' {\it and} `$\overline{-}$', then $\bar{\mathrm{W}}$ measured `$\overline{-}$'.''
Hence, the premise of Statement 1 involves the last term in Eq.(\ref{StateTimeFinal}), 
whereas the conclusion of Statement 1$'$, which is the premise of Statement 2, 
involves both the 3rd and 4th terms. 
A ``shift effect'' is present, and, 
not surprisingly, the logical statement generated using transitivity is ``50\% false''.  
The conclusion of Statement 2, namely that Agent F measured 
the unbarred spin to be up at time $t_2$, 
uses a quantum interference effect similar to the one involved in combining Statements 3 and 4. 
In fact, the ``wavefunction structures'' and the ``If ... then ...'' statements for the two cases are isomorphic.

The ``If ... then ...'' Statements 1 through 4 in the Frauchiger-Renner gedanken experiment 
are all valid in unitary quantum mechanics when considered in isolation. 
In reference \refcite{FRGE}, 
a special run is selected, namely, the one in which agents W and $\bar{\mathrm{W}}$ measure `$-$' and `$\overline{-}$'.
What happens when one considers the effect of imposing this?
The answer is that one is restricting the wavefunction to the last term in Eq.(\ref{StateTimeFinal}), 
and Statements 2, 3 and 4 all become invalid, 
thereby ruining the analysis in reference \refcite{FRGE}. 
For example, the conclusion of Statement 3, which is
Agent $\bar{\mathrm{F}}$ measured the barred spin to be down at time $t_1$, 
is {\it not} a consequence of the premise 
``Agent F measured the unbarred spin to be up at time $t_2$ 
AND $\bar{\mathrm{W}}$ measures `$\overline{-}$' at time $t_3$ 
AND Agent W measures `$-$' at time $t_4$,'' 
as one can verify. 
This is another example of the fact that
the use of certain rules of logic do not always properly take into account 
the combined effects of measurements made by the agents. 
Intuitively, it is easy to understand why Statements 2 and 4 
are rendered invalid when the `$-$' - `$\overline{-}$' condition is imposed: 
The validity of both these statements depends on a perfect quantum interference cancellation between two terms in the wavefunction. 
Anything that disrupts the delicate cancellation will render the corresponding statement false.
Consider Statement 2, for example. 
Its validity depends on a cancellation involving the first and third terms in Eq.(\ref{StateTime3}) 
and the evolution of these two terms going forward in time. 
However, the constraint that agents W and $\bar{\mathrm{W}}$ respectively measure 
`$-$' and `$\overline{-}$' affects all three terms in Eq.(\ref{StateTime3}) and upsets the quantum interference cancellation.

To generate the Contradictory Statemen of reference \refcite{FRGE}, 
the premises of Statements 1 through 4 must be all true at once. 
These premises are ``Agent W measures `$-$' '',  ``$\bar{\mathrm{W}}$ measures `$\overline{-}$' '', etc..
When Agent W measures `$-$' at time $t_4$, 
it can be verified that Agent $\bar{\mathrm{F}}$ had to have measured 
barred spin to be up at time $t_1$. 
The premise of Statement 4 is ``Agent $\bar{\mathrm{F}}$ measured the barred spin to be down at time $t_1$''.
Hence, whenever the premise ``Agent W measures `$-$' at time $t_4$'' is true,  
the premise of Statement 4 is false, and vice versa. 
There are no instances when the premises of Statements 1 through 4 are all true at once.  
So, it is not surprising that, in incorrectly using the rules of logic for statements about measurements 
in their gedanken experiment, 
Frauchiger and Renner can arrive at the logically contradictory statement ``If, at time $t_4$, agents W and $\bar{\mathrm{W}}$ respectively measure 
`$-$' {\it and} `$\overline{-}$', then agent W can deduce that he will measure `$+$'.

There are additional incompatibilities with the premises: 
If the usual rules of logic hold, 
then $(A \Rightarrow B \textrm{ AND } B \Rightarrow C) \Rightarrow (A \textrm{ AND } B) \Rightarrow C$. 
So, it should be true that $(P_1 \textrm{ AND } P_2\ \textrm{AND}\ P_3 \textrm{ AND } P_4) \Rightarrow C_4$, 
where $P_i$ is the premise of Statement $i$ and $C_4$ is the conclusion of Statement 4. 
Hence, the premises must be all true at once to derive the Contradictory Statement. 
However, the situation is even worse: 
The premise of Statement 2 is 
``Agent $\bar{\mathrm{W}}$ measured `$\overline{-}$' at time $t_3$''. 
The premise of Statement 4 is 
``Agent $\bar{\mathrm{F}}$ measured the barred spin to be down at time $t_1$''.
Now from the anlaysis at the end of Section \ref{MeasurementLogic}, 
we know that if Agent $\bar{\mathrm{W}}$ measured `$\overline{-}$' at time $t_3$,  
then the relevant part of the wavefunction 
had to have evolved from something proportional to $\psibar{F}_{\uparrow} - \psibar{F}_{\downarrow}$ 
(which is the same as $\uspinbar_M - \dspinbar_M$)
at time $t_1$.
This means that we are in the situation described in the paragraph 
that contains Eqs.(\ref{ConjunctionIssue0}) and (\ref{ConjunctionIssuet}) 
except that barred spin is involved: 
$P_2$ and $P_4$ are ``conjunctually incompatible'' meaning that  
it is illegitimate to have them appear in the same logical AND statement. 
One of the premises of Statement 1, namely that 
``Agent W measures `$-$' at time $t_4$'', 
is also ``conjunctually incompatible'' with the premised of Statement 3, 
namely that ``Agent F measured the unbarred spin to be up at time $t_2$'' 
for the same reason as the ``barred'' case. 
In a single-photon interferometric setup implementing the scenario of Frauchiger and Renner, 
reference \refcite{Elouard} observed similar compatibility issues. 

From the results in this section, one can see that there are a plethora of errors 
in the analysis of the gedanken experiment in reference \refcite{FRGE}.

\section{The Frauchiger-Renner Wigner/Friend Measurements}
\label{WignerMeasurement}

In this section, we reveal a technical problem with the Wigner/friend measurements used in the Frauchiger-Renner gedanken experiment.
The experiment makes use of two such measurements. 
Suppose that Agent F measures a qubit, which we represent as the spin of a spin-$\oneHalf$ object. 
Let $\Psi^F$ be the wavefunction of Agent F before the measurement is made. 
The wavefunction $\Psi^F$ in general consists of many degrees of freedom -- 
those of the experimentalist and those of her apparatus. 
When Agent F measures the state $\uspin$, 
the wavefunction for F changes: 
at a minimum, the apparatus records the up-spin result 
and the experimentalist notes in her brain that the spin was measured to be up. 
As explained in the Introduction, 
we denote the resulting wavefunction by ${\Psi}^F_\uparrow$. 
If the spin-$\oneHalf$ object ``survives'', its degrees of freedom are included in ${\Psi}^F_\uparrow$.
In cases in which the qubit states are represented by the right and left polarizations of a photon 
and the photon is destroyed during the measurement, 
${\Psi}^F_\uparrow$ does not include the qubit degree of freedom;  
the measurement process is still represented by $\Psi^F \uspin \to {\Psi}^F_\uparrow$. 
When Agent F measures the state $\dspin$, 
statements similar to the above apply and the process is presented by  $\Psi^F \dspin \to {\Psi}^F_\downarrow$.

A Wigner/friend measurement involves a new Agent W who makes a measurement on the 
${\Psi}^F_\uparrow$ and ${\Psi}^F_\downarrow$ states. 
In the Frauchiger-Renner gedanken experiment, 
Agent W makes the measurement in the basis ${\Psi}^F_+ = ({\Psi}^F_\uparrow + {\Psi}^F_\downarrow)/\sqrt{2}$ 
and ${\Psi}^F_- =  ({\Psi}^F_\uparrow -  {\Psi}^F_\downarrow)/\sqrt{2}$.
If  $\Psi^W$ is the wavefunction before the ``Wigner'' measurement is made, then, as in the case of Agent F above, 
$\Psi^W$ is affected by the measurement and the process is represented by 
$\Psi^W {\Psi}^F_+ \to {\Psi}^W_+$ and $\Psi^W {\Psi}^F_- \to {\Psi}^W_-$.
The degrees of freedom of F are included in ${\Psi}^W_+$ and ${\Psi}^W_-$. 

In the Frauchiger-Renner gedanken experiment, 
Agent W has an almost impossible task in measuring the `$+$' and `$-$' states of such a complicated system,\cite{Elouard}  
and as L\'idia del Rio and Renato Renner note     
the Wigner agents must ``have excellent quantum control of other agents’ memories and labs''\cite{delRioRenner} 
in order to conduct their measurements on agents F and $\bar{\mathrm{F}}$, 
at lease in the way that the Wigner measurements are presented in reference \refcite{FRGE}.
However, there is also a tremendous burden on Agent F (and Agent $\bar{\mathrm{F}}$). 
If Agent F is a complicated object -- and indeed up until now we have been assuming this 
since F consists of the experimentalist and her equipment -- 
then it is unlikely that the same ${\Psi}^F_\uparrow$ is produced each time $\uspin$ is measured. 
This is a problem for Agent W because in the Wigner/friend experiment he is not allowed 
to examine ${\Psi}^F_\uparrow$ and ${\Psi}^F_\downarrow$.
So, how can he measure $({\Psi}^F_\uparrow + {\Psi}^F_\downarrow)/\sqrt{2}$ and 
$({\Psi}^F_\uparrow - {\Psi}^F_\downarrow)/\sqrt{2}$ if he does not know what these states are? 
The solution is that Agent F must respond to the measurement of the spin in a predetermined known way 
to produce specific ${\Psi}^F_\uparrow$ and ${\Psi}^F_\downarrow$, and Agent W must be informed of these ``known'' states 
at the start of the experiment and before the measurements are made. 
It is clearly impossible for Agent F to produce a specified ${\Psi}^F_\uparrow$ or ${\Psi}^F_\downarrow$ 
given that Agent F is such a complicated object. 
Among things, the center of mass coordinate of Agent F is a continuous variable that cannot be precisely fixed 
because of the Heisenberg uncertainty principle.  

To shed some light on the issue, 
consider replacing all the quantum degrees of freedom associated with Agent F 
with a system of 100 qubits, with each qubit having two states: up and down. 
During the measuring process, these 100 qubits are ``disturbed'' randomly but the signal 
of the experimental outcome is encoded in the last qubit: 
If the spin was up (respectively, down), then 
the last qubit is up  (respectively, down) after the measurement is performed. 
Now Agent F and W agree, for example, that ${\Psi}^F_\uparrow$ (respectively,  ${\Psi}^F_\downarrow$) 
corresponds to all the qubits being up (respectively, down). 
Now when the experiment takes place Agent F has the very difficult task of controlling how 
the experiment effects the first 99 bits. 
It is very unlikely that the qubits will all be up or all be down. 
Hence, almost all the time, Agent W gets no signal when he tries to make his measurement. 
So, the ``100-spin case'' is difficult but still doable in principle. 
However, the situation is rendered impossible when one considers that 
among the enormous number of quantum degrees of freedom of Agent F,  
there are many -- in fact most -- which are continuous and not discrete. 

A way around this problem is to have Agent W interact only with a small ``important'' subset of the degrees of freedom of F. 
These degrees of freedom might include a qubit in a data base that recorded the reading as up or down (as in the previous paragraph) 
as well as data bytes providing the time of the measurement $t_m$ and 
statements such as ``Agent F knows that the measurement was up at time $t_m$'' 
that are used in reference \refcite{FRGE}, and so on. 
In other words, if we write $\Psi^F = \Psi'^F \ket{S}$ where \ket{S} indicates a state associated 
with these ``important'' degrees of freedom, then we can have 
$\Psi^F  \ket{\uparrow} =  \Psi'^F \ket{S} \ket{\uparrow} \to \Psi'^F  \ket{S}' \ket{\uparrow}_M$, 
where $\ket{\uparrow}_M$ indicates a specific state for the up case that provides the ``measurement recording information.'' 
For the down spin case, $\Psi^F  \ket{\downarrow} \to \Psi'^F  \ket{S}' \ket{\downarrow}_M$, 
where, again, $\ket{\downarrow}_M$ is some specific state. 
Agent W can then be supplied in advance with $\ket{\uparrow}_M$ and $\ket{\downarrow}_M$. 
One simple realization of this is $\ket{S} =  \ket{\uparrow}_M$. 
One can then have $\ket{\uparrow}_M \ket{\uparrow} \to \ket{\uparrow}_M  \ket{\uparrow}$ ($= \ket{\uparrow} \ket{\uparrow}_M$) 
and  $\ket{\uparrow}_M \ket{\downarrow} \to \ket{\downarrow}_M \ket{\uparrow}$  ($= \ket{\uparrow} \ket{\downarrow}_M$)  
(which is just spin exchange) as the measurement process, 
in which case $\ket{S}' = \ket{\uparrow}$. 

If the initial state is $\ket{\uparrow}$ for example, 
then $\Psi^W {\Psi}^F \ket{\uparrow} = \Psi^W \Psi'^F \ket{S} \ket{\uparrow} \to \Psi^W \Psi'^F \ket{S}' \ket{\uparrow}_M = 
( \Psi'^F  \ket{S}' /  \sqrt{2}) \Psi^W ( (\ket{\uparrow}_M + \ket{\downarrow}_M)/ \sqrt{2} +  (\ket{\uparrow}_M - \ket{\downarrow}_M) \sqrt{2} ) = 
\Psi'^F  \ket{S}' (\Psi^W_+ + \Psi^W_-)  / \sqrt{2}$.
Likewise, $\Psi^W {\Psi}^F \ket{\downarrow} \to \Psi'^F \ket{S}' (\Psi^W_+ - \Psi^W_-)  / \sqrt{2}$.
It is easy to see that $\Psi''^F = \Psi'^F \ket{S}' $ plays no role in the Wigner/friend experiment: 
It is just an overall factor in the wavefunction 
from the time of the measurement by Agent F henceforth, and therefore can be ignored in the analysis. 
When this is done, $\ket{\uparrow}_M $ and $\ket{\downarrow}_M$ act like a qubit but with messages associated with them.
This simplification justifies, for example, the replacement of a friend agent by photon polarizations 
as is done in reference \refcite{WignerExperiment}.
Although there are many possibilities for agents F and W to agree in advance 
on what  ${\Psi}^F_\uparrow$ and ${\Psi}^F_\downarrow$ are,
the result that these two states need to be replaced 
by two specific states, which we can call $\ket{\uparrow}_M$ and $\ket{\downarrow}_M$, is a general result. 
Agent W can still involve the many degrees of freedom of the human experimentalist and his equipment. 

Note, when the procedure described in the previous two paragraphs is used, 
that the final wavefunction for the experimentalist must be the same whether 
the initial spin state is $\ket{\uparrow}$ or $\ket{\downarrow}$.
This means that the experimentalist cannot be conscious of the experimental outcome. 
The definition of a {\it measurement} on a quantum system is not universally defined in physics. 
In its definition, one might require a human or intelligent being to be conscious of the outcome, 
in which case, a Wigner measurement of the type occurring in the Frauchiger and Renner gedanken experiment is impossible, 
given the results in the first four paragraphs of this section. 
Alternatively, the definition of a quantum measurement might only require the outcome to be ``recorded'' or ``registered'', 
which means that it is not necessary for an intelligent being to be aware the experimental result  
(especially if it is potentially possible to verify the recorded result by human participation in the manner used at the end of Section \ref{WavefunctionLogic}).
In this case, a Wigner measurement is possible but it must be made on an entity 
without a center of mass degree of freedom such as a spin, a photon polarization, a tensor product of these, et cetera. 
The Frauchiger and Renner experiment is therefore only possible 
if the states $\ket{\uparrow}_M$, $\ket{\downarrow}_M$, $\uspinbar_M$ and $\dspinbar_M$ 
are of this form. 
However, such states are necessarily microscopic. 
Hence, when the Wigner agents make their measurements,
it is on microscopic entities, in which case, they are ``ordinary'' quantum measurements. 
Regardless of the problems with the transitive property of logic, 
{\it the Frauchiger-Renner gedanken experiment cannot be making a statement about a macroscopic system.} 

If the measurements by Agents F and $\bar{\mathrm{F}}$ are not considered measurements but recordings, 
then the subscripts `M' in Sections \ref{TheWavefunctions} through \ref{IssuesWithLogic} are misleading 
and should be replaced by `R'. 
For completeness, we provide a brief description on how the Frauchiger-Renner gedanken experiment 
is modified to take this into account. 
One needs to avoid saying that Agent $\bar{\mathrm{F}}$ and Agent F make measurements on the barred and unbarred spins, 
since almost all but a few discrete quantum degrees of freedom of these two agents are affected. 
For example, the original first step, which is  
``Agent $\bar{\mathrm{F}}$ measures the spin of the barred spin-$\oneHalf$ object in the $z$-direction at time $t_1$,'' 
needs to be replaced by 
``An experimental procedure on the barred spin-$\oneHalf$ object in the $z$-direction at time $t_1$ is performed 
and the outcome is recorded in another spin-$\oneHalf$ object as $\uspinbar_R$ or $\dspinbar_R$.'' 
A similar replacement occurs for step two. 
The subscript ``M'' is replaced by ``R'' on barred and unbarred spins 
in Sections \ref{TheWavefunctions} and \ref{TheArgument}.
Statement 1 and Statement 1$'$ are unchanged but Statements 2 to 4 become: 

Statement 2: If Agent $\bar{\mathrm{W}}$ measured `$\overline{-}$' at time $t_3$, then 
the unbarred spin was recorded to be up at time $t_2$. 

Statement 3: If the unbarred spin was recorded to be up at time $t_2$,  
then the barred spin was recorded to be down at time $t_1$.

Statement 4: If the barred spin was recorded to be down at time $t_1$, then 
Agent W will measure `$+$' at time $t_4$. 
\newline
If these four statements could be combined using the transitive property of logic, 
then one would still obtain the Contradictory Statement of Section \ref{TheArgument}. 

It should be clear that any Wigner measurement on a linear combination of 
Agent F states involving {\it all the degrees of freedom of a human experimentalist and her equipment} is, in general, impossible. 
For the case in which Agent F performs a measurement on a spin-$\oneHalf$ object in the $z$-direction, 
Agent W is unable to perform a measurement using a basis of 
$\cos{\theta} {\Psi}^F_\uparrow + \sin{\theta} {\Psi}^F_\downarrow$ 
and $\sin{\theta} {\Psi}^F_\uparrow - \cos{\theta} {\Psi}^F_\downarrow$ 
for any $\theta$ for which both $\cos{\theta} \ne 0$ and $\sin{\theta} \ne 0$.
Note that $\theta = \pi/4$ is the Frauchiger-Renner case.
Hence, the only basis in which a Wigner agent can make a Wigner measurement on Agent F in the gedanken experiment is 
one that is ``aligned'' with the basis that Agent F used. 
For the case of the spin-$\oneHalf$ object discussed in this section, 
the basis is
${\Psi}^F_\uparrow$ and ${\Psi}^F_\downarrow$, 
and the measurement process is quite simple for this: 
Agent W can meet with Agent F and just ask her what she measured. 
The process is schematically represented by 
$\Psi^W {\Psi}^F_\uparrow \to {\Psi}^W_\uparrow$ and $\Psi^W {\Psi}^F_\downarrow \to {\Psi}^W_\downarrow$.

\section{Discussion and Conclusions}
\label{conclusion}

In their work, Frauchiger and Renner concluded that 
quantum theory cannot be extrapolated to complex systems in a straightforward manner. 
However, the generation of a contradiction arises 
only if Assumption (S) is replaced by the stronger Assumption (L),\cite{RennerEmailA}
which states that  statements by agents concerning measurements of wavefunctions 
obey standard rules of logic. 
One might naively think that Assumption (L) should also hold.\cite{RennerEmailB}
However, we have shown in this paper that this is not the case: 
Statements about measurements cannot necessarily be combined 
to generate new statements using the standard rules of classical logic. 
Once one understands this, 
there is nothing, in principle, ruling out quantum mechanics 
being able to govern complex and macroscopic systems. 
Unitary quantum mechanics does not have an Einstein-Podolsky-Rosen paradox\cite{epr} 
as was pointed out by Hugh Everett in his Ph.D. thesis\cite{EverettThesisA,EverettThesisB}, 
nor does it have a measurement problem;\cite{SSMeasurement} 
indeed, the wavefunctions in Equations (\ref{DeterminedMeasurement}) and (\ref{UncertainMeasurement}) 
correctly describe both the microscopic and macroscopic situations,  
so there is a consistent transition in going from small scales to large scales without wavefunction collapse, 
implying that there is no Heisenberg cut\cite{HeisenbergCut} 
in contradiction to what references \refcite{delRio} and \refcite{delRioRenner} believe. 
If a Heisenberg cut existed, then it would undermine the conclusion of Frauchiger and Renner
because classical mechanics replaces certain aspects of unitary quantum mechanics above the cut 
thereby ``not allowing'' quantum mechanics to fully operate in the macroscopic world. 

Section \ref{IssuesWithLogic} shows that by {\it explicit calculation} that the use of logical transitivity 
in the Frauchiger-Renner gedanken experiment is violated in three separate instances. 
This result has been verified by three authors\footnote{While waiting for the editorial 
decision of the submission of my work to Nature Communications, 
I asked some researchers to determine the consequence of the measurement of Agent F 
on the measurement of Agent W.}
 for the case of Bohmian Mechanics\cite{BohmianMechanics} 
(which a modification of quantum mechanics involving a tracking field that maintains unitarity)
for combining statements 3 and 4.\cite{DustinLazarovici,AlexMatzkin,DmitriSokolovsk}
In Section  \ref{IssuesWithLogic}, 
we also provide three explanations for why statements in the Frauchiger-Renner gedanken experiment 
cannot be combined using logic: \\
(i) Combining Statements 1$'$ and 2, Statements 2 and 3, as well as Statements 3 and 4 
involve a ``shift effect'' (See Section \ref{MeasurementLogic}), 
and this invalidates the use of transitivity for these pairs of statements. 
Eq.(\ref{ViolationFraction}) provides a quantification of the violation, 
and it is 50\% for each of the above three uses of transitivity. \\
(ii) The premise of the first ``If ... then ..." statement is false in a certain fraction of the instances 
for which the premise of the second ``If ... then ..." statement is true. 
This fraction coincides with the result in Eq.(\ref{ViolationFraction}).
When the premise of the first statement is false while the premise of the second statement is true, 
a false result is generated from the ``If ... then ..." statement obtained by combining the two ``If ... then ..." statements 
using transitivity.
This happens in combining Statements 1$'$ and 2, Statements 2 and 3, and Statements 3 and 4. \\
(iii) Combining Statements 1$'$ and 2 and Statements 3 and 4 involve a quantum interference effect. 
This interference effect is upset by the measurement associated with the premise of the first statement. 
In other words, the ``If ... then ..." statement obtained by using transitivity does not properly take into account 
the effect of the measurement perform by the first agent on the measurement performed by the second agent. 

There are five problems with the analysis performed by Frauchiger and Renner. \\
(1) The transitive property of logic is not valid in combining pairwise the four statements to generate a contradiction. 
This is related to another issue: \\
(2) In logic, $(A \Rightarrow B \textrm{ AND } B \Rightarrow C) \Rightarrow ((A \textrm{ AND } B) \Rightarrow C)$.
However, in unitary quantum mechanics for the cases involving Statements 1 - 4, 
$(A \textrm{ AND } B) \Rightarrow C'$, where $C'$ is a conclusion that is different from $C$. 
This shows that this rule of logic is violated in the Frauchiger-Renner gedanken experiment.  
This is relevant for the generation of the contradictory statement because it is unclear 
whether one should use $C$ or $C'$; 
the argument in the Frauchiger-Renner publication needs to use $C$ 
to generate the contradictory statement. 
However, in combining pairs of statements $C'$ turns out to be the correct conclusion. 
In short,  Frauchiger and Renner used the wrong formula for combining the statements.\\
(3) Frauchiger and Renner run their experiment until both agents W and  $\bar{\mathrm{W}}$ respectively
measure `$-$' and `$\overline{-}$'.
It can be shown that restricting the run to this case 
renders Statements 2, 3 and 4 false.  \\
(4) Instead of performing runs until agents W and  $\bar{\mathrm{W}}$ respectively measure `$-$' and `$\overline{-}$',
one can replace Statement 1 by Statement 1$'$ (See Section \ref{TheArgument}). 
All premises must be true at once to obtain a valid Contradictory Statement. 
One of these premises of Statement 1$'$ is that Agent W measures `$-$'. 
The premise of Statement 4 is that Agent $\bar{\mathrm{F}}$ measures the barred spin to be down. 
Section \ref{IssuesWithLogic} showed that when the premise that Agent W measures `$-$' is true, the premise of Statement 4 is false,
and when the premise of Statement 4 is true, the premise that Agent W measures `$-$' is false. \\
(5) If one formulates the generation of the contradictory statement as
$(P_1 \textrm{ AND } P_2\ \textrm{AND}\ P_3 \textrm{ AND } P_4) \Rightarrow C_4$, 
where $P_i$ is the premise of Statement $i$ and $C_4$ is the conclusion of Statement 4, 
then one finds that $P_2$ and $P_4$ are incompatible for use with logical conjunction. 
In addition,  $P'_1$ and $P_3$ are incompatible.
See the next-to-the-last paragraph of Section \ref{IssuesWithLogic}.

In the above, (1) and (2) involve the issue of combining two successive statements using transitivity in the Frauchiger-Renner gedanken experiment. 
Items (3), (4), and (5) point out other problems in combining the four statement to produce the contradictory statement. 
In short, the usual rules of logic cannot be used on Statements 1 - 4 in the Frauchiger-Renner gedanken experiment. 

At a minimum, our work has cleared up a misconception created by reference \refcite{FRGE} 
that has already reached mainstream scientific media.\cite{QuantumMagazine,ScientificAmerican,WikipediaWigner,ScienceDaily,ScienceAAAS,SabineHossenfelder}
We have also obtained other important results. 
We developed the concept of quantum logic and used it 
to deduce physical and mathematical consequences from knowledge of a wavefunction. 
We have learned that one must be careful in using many of the`` standard'' rules of classical logic 
for statements about wavefunctions and measurements. 
Sections \ref{WavefunctionLogic} - \ref{IssuesWithLogic} shed light on why the violations of the rules of logic 
are expected in certain circumstances.
In Section \ref{WignerMeasurement}, we pointed out a restriction on Wigner/friend experiments. 
If this restriction is imposed, 
then the Wigner/friend measurement of the Frauchiger-Renner gedanken experiment 
becomes an ordinary quantum measurement, 
allowing the possibility of carrying out the experiment in a real laboratory setting. 

The Frauchiger-Renner gedanken experiment does not demonstrate problems with quantum mechanics at macroscopic scales, 
but it is a useful laboratory for exploring quantum mechanics, quantum logic, and Wigner/friend measurements. 

\section*{Acknowledgments}

I thank the members (and, in particular, Alexios Polychronakos) 
of the theory groups at City College (CCNY) and City University of New York (CUNY) for feedback. 
I thank Renato Renner for a series of interesting email exchanges that 
helped the author uncover items (ii), (iii), and (4) in the Discussion and Conclusions Section. 

\appendix

\section{The Frauchiger-Renner Assumptions in Unitary Quantum Mechanics}
\label{ThreeAssumptionInUQM}

In this appendix, we discuss some issues with Assumptions (S) and (C). 
These problems are not ``fatal'' in the sense that they can be replaced by other assumptions. 


Assumption (S), as formulated in reference \refcite{FRGE},  
says that ``if Agent A is certain of the statement $x = v$ at time $t$,  
then he has to deny the statement $x \ne v$ at time $t$.'' 
In unitary quantum mechanics, an agent can only be certain of the outcome of a measurement 
when wavefunction collapse is not needed to explain it. 
For example, if Agent A was trying to determine if a spin was up or down and the initial state $S_0$ to be measured
was of the form $S'_0 \uspin$ where $S'_0$ is a wavefunction not involving the spin degree of freedom 
(we also include the case in which $S'_0$ is a phase), 
then this is similar to case (i) in Eq.(\ref{DeterminedMeasurement}): 
After the measurement, the wavefunction is $S'_0 \Psi_{\uparrow}^A$ 
and $\Psi_{\uparrow}^A$ embodies the statement that Agent A measured the spin to be up and knows that it is up. 
He must deny that it is down in agreement with Assumption (S). 
On the other hand, if Agent A was trying to determine if a spin was up and the initial state $S_0$ 
was of the form $a_{\uparrow} \uspin + a_{\downarrow} \dspin$,  
where now $a_{\uparrow}$ and $a_{\downarrow}$ can be wavefunctions not involving the spin, 
then the situation is similar to case (ii) of the Introduction: 
After the measurement, the wavefunction is the result in Eq.(\ref{UncertainMeasurement}) 
and, as explained in the Introduction when discussing the Measurement Rule, 
both $\Psi_{\uparrow}^A$ and $\Psi_{\downarrow}^A$
embody the concept that Agent A cannot be sure that the spin is up or down with certainty 
even though an output from an experimental device seems to be indicating a definite result for the spin state 
in $\Psi_{\uparrow}^A$ and in $\Psi_{\downarrow}^A$.
After the measurement is made, 
the wavefunction associated with Agent A in case (ii) 
involves contributions from two different distributions (of Agent A's quantum constituents):  
one associated with $\Psi_{\uparrow}^A$ and one associated with $\Psi_{\downarrow}^A$.
These two embody different ``statements'' 
but common to both is the lack of certainty about the measurement 
when the initial state is $a_{\uparrow} \uspin + a_{\downarrow} \dspin$.
Basically, the situation is that the wavefunction for the quantum degrees of freedom in Agent A is in a linear superposition, 
and as such, there is uncertainty. 
So, for case (ii), Assumption (S) is satisfied because the premise (``Agent A is certain of the statement $x = v$ at time $t$'')
is not satisfied.
Alternatively, since the statement ``Agent A is certain of the statement $x = v$ at time $t$'' 
is a ``classical'' one but the situation is a ``quantum'' one,
the premise of Assumption (S) can be considered as not making sense 
when Agent A is in a linear superposition as happens in case (ii).

Renner and Frauchiger use Assumption (S) in the following way: 
Four statements about wavefunction measurements are established 
in the following form: premise$_i$ implies conclusion$_i$ (where $i =$ 1, 2, 3 or 4), 
and where conclusion$_i$ =  premise$_{i+1}$, for $i =$ 1, 2, and 3).
When these statements are combined assuming that the transitive property of logic is valid, they yield premise$_1$ implies conclusion$_4$.
Premise$_1$ contains a statement of the form ``Agent A measures $x$ to be $v$ at time $t$'' 
and conclusion$_4$ contains a statement of the form ``Agent A measures $x$ to be $v'$ at time $t$'' 
where $v' \ne v$.{\footnote{In the Wigner/friend gedanken experiment, 
Agent A is Agent W, $x$ is unbarred measured spin, $v = \ket{-}_M$, $v' =\ket{+}_M$ and $t = t_4$.}} 
So, being cautious, Renner and Frauchiger require Assumption (S) so that premise$_1$ and conclusion$_4$ produce the Contradictory Statement. 
In unitary quantum mechanics, 
the above same four statements are derivable. 
However, the problem is not that premise$_1$ implies conclusion$_4$ is a contradiction; 
the problem is that the transitive property of logic does not always apply when combining statements about wavefunction measurements. 

Assumption (S) is also inconsistent with the experimental procedure of the Frauchiger-Renner gedanken experiment. 
Agent W, for example, is supposed to perform a measurement 
using the basis $\{ (\psi^{F}_{\uparrow} + \psi^{F}_{\downarrow})/\sqrt{2}$,  $ (\psi^{F}_{\uparrow} - \psi^{F}_{\downarrow})/\sqrt{2}\}$.\footnote{Recall that  
$\psi^{F}_{\uparrow} = \uspin_M$ (respectively $\psi^{F}_{\downarrow} = \dspin_M$), 
is a wavefunction for Agent F in which unbarred spin is measured to be up (respectively, down).
}
Since this is a Schr\"odinger Cat state involving Agent F measuring spin up and Agent F measuring spin down,
it is inconsistent with the assumption that an agent be certain of a measurement. 

A careful reading of reference \refcite{FRGE} reveals that, in the ``proof'', 
one only needed Assumption (S$'$): If a logical contradiction arises, one can conclude that something is wrong. 
In any case, Assumption (S) needs to be replaced by Assumption (L). 

Now consider Assumption (C).
It says,\\ 
\noindent
Suppose that Agent A has established that

``I am certain that Agent A$'$, upon reasoning within the same theory as the one I am using, is certain that $x = \xi$ at time $t$.''\\
Then Agent A can conclude that

``I am certain that  $x = \xi$ at time $t$.''\\
\linebreak
\noindent
The problem with this is that in unitary quantum mechanics, 
one can be sure of a measurement result only if the situation is as in Eq.(\ref{DeterminedMeasurement}).
However, the situation for the Frauchiger-Renner gedanken experiment involves Eq.(\ref{UncertainMeasurement}).
So, the premise of Assumption (C) is always false because of the word ``certain'' in it. 

The problems with assumptions (S) and (C) have a common origin: 
trying to force ``classical thinking'' onto a quantum situation. 

Frauchiger and Renner used (C) and the agreement among agents as to which quantum theory to use
(which because of Assumption (U) must be unitary quantum mechanics) 
to deduce Statements 1 - 4. 
However, in unitary quantum mechanics, Statements 1 - 4 can be deduced 
from the description of the gedanken experiment. 
Hence, Assumption (C) is not needed.\footnote{In the context of the Frauchiger-Renner Gedanken experiment,
this is good thing since we have just shown that  Assumption (C) has problems.
}
In the rest of the Appendix, we explain why.

In unitary quantum mechanics for the extended Wigner/friend gedanken experiment, 
all agents know the initial spin part of the wavefunction and the experimental procedure.
From this information, 
agents, an outside observer, or even the reader of this article, 
can deduce the overall structure of the wavefunction at various times. 
For example, 
at time $t_4$, 
Agent $\bar{\mathrm{W}}$ does not know that $\Psi_4$ is precisely as in Eq.(\ref{StateTimeFinal}), 
but he does know that the structure is 
\begin{equation}
     \Psi_4 = \frac{1}{\sqrt{12}}  \big(  
       3 {\phibar{W}_+} {\Phi^W_+} - \phibar{W}_+ {\Phi^W_-} - \phibar{W}_- {\Phi^W_+} - \phibar{W}_- {\Phi^W_-}
   \big)
\ , 
\label{StateTimeFinalStructure} 
\end{equation}
where the $\Phi_a^A$ (with $a =  +$ or $-$ and $A = W$ or $\bar{W}$) embody the same messages and statements 
(given at the beginning of Section \ref{TheArgument}) as the $\Psi_a^A$.

In the case of the wavefunction at time $t_2$, 
Agent W actually knows the spin part is of the form 
\begin{equation}
          \frac{(\uspinbar_M' \dspin_M + \dspinbar_M' \uspin_M + \dspinbar_M' \dspin_M)}{\sqrt{3}} 
\ ,
\label{StateTime2Structure} 
\end{equation}
where $\uspinbar_M'$ and $\dspinbar_M'$ embody 
the same messages and statements as $\uspinbar_M$ and $\dspinbar_M$.
He knows that the $\uspin_M$ and $\dspin_M$ in Eq.(\ref{StateTime2Structure}) 
are the same as the states in Eq.(\ref{StateTime2}) because 
he and Agent F have a prior agreement as to what the measured states will be, 
as explained in Section \ref{WignerMeasurement}. 

Once the structure of the wavefunction is known, 
the methods at the end of Section \ref{TheArgument} allow anybody to derive Statements 1 - 4. 
In reference \refcite{FRGE}, the statements are expressed somewhat differently than in Section \ref{TheArgument}.
For example, Statement 4 in reference \refcite{FRGE} is: 

Statement 4: If I, Agent $\bar{\mathrm{F}}$, measure the barred spin to be down at time $t_1$, then 
I am certain that Agent W will observe `$+$' at time $t_4$. \\
In unitary quantum mechanics, there is no need to phrase statements with terms like such as ``I,'' and ``I am certain that'' 
because anyone can deduce that If Agent $\bar{\mathrm{F}}$ measures the barred spin to be down at time $t_1$, 
then Agent W will observe `$+$' at time $t_4$.
Knowing the initial state, and who measured what, at what time, and in what manner is sufficient to derive Statements 1 - 4. 
Unitary quantum mechanics is ``powerful'' in this regard. 
Indeed, in the paragraphs containing Eqs.(\ref{StateTime4A}) and (\ref{StateTime4B}), 
we showed that knowledge of the final state and the experimental procedure is also sufficient to derive Statements 1 - 4, 
a result that many might find surprising.



\begin{thebibliography}{99}

\bibitem{FRGE} D. Frauchiger and R. Renner, 
Quantum theory cannot consistently describe the use of itself, 
{\it Nature Communications} {\bf 9} (2018) 3711.

\bibitem{WignersFriendA}
E. P. Wigner, ``Remarks on the mind-body question'', in I. J. Good, \\ 
{\it The Scientist Speculates}, (Heinemann, London, 1961).

\bibitem{WignersFriendB}
D. Deutsch, Quantum theory as a universal physical theory, \\
{\it Int. J. of Theo. Phys.} \textbf{24} (1985) 1--41. 

\bibitem{delRio}
N. Nurgalieva and L. del Rio,
Inadequacy of modal logic in quantum settings, 
{\it Electronic Proceedings in Theoretical Computer Science}, {\bf 287} (2019) 267 -- 297.

\bibitem{RennerEmailA}
Renato Renner, private communication.\\
``I would agree that (L) is an assumption that we implicitly made in our paper'', 
email correspondence received from Renato Renner on 3/17/22;\\
``I agree that we are using the transitivity of logic in our proof --- for example to derive these further implied statements.'', 
email correspondence from Renato Renner on 5/21/22.

\bibitem{RennerEmailB}
Renato Renner, private communication.\\
``I would argue that the applicability of logical reasoning is the basis of any scientific argument'',
email correspondence received from Renato Renner on 3/12/22; \\
``I find it hard to imagine how to do science in a world in which this [the violation of logic] is the case'', 
email correspondence from Renato Renner on 3/17/22;\\
``I agree with your claim that we implicitly assumed the transitivity of logic for our argument. 
However, I disagree with your conclusion that this assumption is problematic.'',
email correspondence from Renato Renner on 5/1/22.

\bibitem{LazaroviciHubert}
D. Lazarovici and M. Hubert, How Quantum Mechanics can consistently describe the use of itself,
{\it Sci. Rep.} {\bf 9} (2019) 470.

\bibitem{SudberyA}
A. Sudbery, Single-World Theory of the Extended Wigner's Friend Experiment, 
{\it Found. Phys} {\bf 47} (2017) 658--669. 

\bibitem{SudberyB}
A. Sudbery, The hidden assumptions of Frauchiger and Renner, \\
{\it Inter. J. of Quant. Found.} {\bf 5} (2019) 98--109.

\bibitem{Yang}
J. M. Yang, Consistent Descriptions of Quantum Measurement, \\
{\it Found. Phys.}  {\bf 49},  (2019) 1306--1324.

\bibitem{LoeligerVontobel}
H.-A. Loeliger and P. O. Vontobel, Quantum Measurement as Marginalization and Nested Quantum Systems, 
{\it IEEE Trans. on Info. Theory.}  {\bf 66} (2020)  3485--3499.

\bibitem{MatzkinSokolovskiA}
A. Matzkin and D. Sokolovski, Wigner-friend scenarios with noninvasive weak measurements, 
Phys. Rev. {\bf A 102} (2020) 062204. 

\bibitem{MatzkinSokolovskiB}
D. Sokolovski and A. Matzkin, Wigner’s Friend Scenarios and the Internal Consistency of Standard Quantum Mechanics, 
Entropy {\bf 23} (2021) 1186. 

\bibitem{WaaijerNeerven}
M. Waaijer and J.v. Neerven, 
Relational Analysis of the Frauchiger–Renner Paradox and Interaction-Free Detection of Records from the Past,
{\it Found. Phys.}  {\bf 51} (2021) 45.

\bibitem{Elouard}
C. Elouard, P. Lewalle, S. K. Manikandan, S. Rogers, A. Frank, and A. N. Jordan, 
Quantum erasing the memory of Wigner's friend,
{\it Quantum}  {\bf 5} (2021) 498. 

\bibitem{SSMeasurement} S. Samuel, 
See Section 6 of On Wavefunction Collapse, the Einstein-Poldolsky-Rosen Paradox and Measurement 
in Quantum Mechanics and Field Theory, https://arxiv.org/abs/1910.11134 (July, 2021); \\
Also see: The Resolution of the Measurement Problem in Unitary Quantum Field Theory, 
Colloquium given to the Physics Department of the City College of New York on 8/25/21. 
A video of this colloquium is available at https://www.youtube.com/watch?v=PGjeMo03Nco\&t=117s (Accessed on 9/10/23). 

\bibitem{EverettThesisA} 
H. Everett, Theory of the Universal Wavefunction, Thesis, Princeton University, (1956), p. 1--140. 

\bibitem{EverettThesisB} 
H. Everett, `Relative State' Formulation of Quantum Mechanics, \\ 
{\it Reviews of Modern Physics} \textbf{29} (1957) 454--462. 

\bibitem{FuzzyLogin} Fuzzy Logic, https://en.wikipedia.org/wiki/Fuzzy\_logic\  (Accessed on 9/10/23).

\bibitem{delRioRenner}
L. del Rio and R. Renner,
Not even subjective truth survives contact with quantum theory, 
submitted to the Nature journal {\it Matters Arising} in response to the submission of my work to Nature Communications
(unpublished). 

\bibitem{WignerExperiment} M. Proietti, A. Pickston, P. Barrow, D. Kundys, C. Branciard, M. Ringbauer, and A. Fedrizzi, 
Experimental test of local observer independence, Science Advances, \textbf{5}, \\ 
No. 9, https://www.science.org/doi/10.1126/sciadv.aaw9832\  (Accessed on 9/10/23).

\bibitem{epr} A. Einstein, B. Podolsky, and N. Rosen, 
Can Quantum-Mechanical Description of Physical Reality Be Considered Complete?, 
{\it Phys. Rev.} {\bf 47} (1935) 777--780.

\bibitem{HeisenbergCut}
https://en.wikipedia.org/wiki/Heisenberg\_cut\ (Accessed on 9/10/23).

\bibitem{BohmianMechanics}
D. A. Bohm, 
Suggested Interpretation of the Quantum Theory in Terms of`'Hidden Variables' I, 
Physical Review \textbf{85} (2) (1952) 16--179.

\bibitem{DustinLazarovici}
Dustin Lazarovici, private communication.\\
D. Lazarovici verfied that if Agent F measured unbarred spin up at time $t_2$, 
then there is a 50\% chance that Agent W will obtain `$+$' for his measurement and not 100\%. 
The use of logical transitivity by Frauchiger and Renner produces the incorrect 100\% result.,
email correspondence received on 12/7/22.

\bibitem{AlexMatzkin}
Alex Matzkin, private communication.\\
A. Matzkini stated that if Agent F measured unbarred spin up at time $t_2$, 
then there is a 50\% chance that Agent W will obtain `$+$' for his measurement and not 100\%,
email correspondence received on 12/28/22. 

\bibitem{DmitriSokolovsk}
Dmitri Sokolovsk, private communication.\\
D. Sokolovsk confirmed that if Agent F measured unbarred spin up at time $t_2$, 
then there is a 50\% chance that Agent W will obtain `$+$',
email correspondence received on 1/14/23. 

\bibitem{QuantumMagazine} A. Ananthaswamy, New Quantum Paradox Clarifies Where Our Views of Reality Go \\ 
Wrong, 
Quantum Magazine (December 3, 2018), \\ 
https://www.quantamagazine.org/frauchiger-renner-paradox-clarifies-where-our-\\ 
views-of-reality-go-wrong-20181203/\  (Accessed on 9/10/23).

\bibitem{ScientificAmerican} Z. Merali, This Twist on Schr\"odinger's Cat Paradox Has Major Implications for \\ 
Quantum Theory, Scientific American (August 17, 2020), \\
https://www.scientificamerican.com/article/this-twist-on-schroedingers-cat-paradox-\\ 
has-major-implications-for-quantum-theory/\ (Accessed on 9/10/23).

\bibitem{WikipediaWigner} Wikipedia, Wigner's Friend,  
https://en.wikipedia.org/wiki/Wigner\%27s\_friend,  \\ 
See the section on ``An extension of the Wigner's friend experiment'' \\
(Accessed on 9/10/23). 

\bibitem{ScienceDaily}
Searching for errors in the quantum world, Science Daily (September 18, 2018). 
https://www.sciencedaily.com/releases/2018/09/180918114438.htm \\
(Accessed on 9/10/23). 

\bibitem{ScienceAAAS}
Quantum paradox points to shaky foundations of reality, 
Science (of the American Association for the Advancement of Science) (August 17, 2020). \\
https://www.science.org/content/article/quantum-paradox-points-shaky-foundations-reality\ (Accessed on 9/10/23). 

\bibitem{SabineHossenfelder}
Has quantum mechanics proved that reality does not exist?, \\
Youtube video by Sabin Hossenfelder with over 730,000 views as of 9/10/23. \\
https://www.youtube.com/watch?v=Wsjgtp9XZxo\&t=406s (Accessed on 9/10/23).

\end{thebibliography}
\end{document}